\documentclass[notitlepage,11pt,
amsmath,preprintnumbers,superscriptaddress,nofootinbib,aps] 
{revtex4-1} 

\usepackage{amssymb,url,bm}

\usepackage{amssymb,amsfonts,multirow}

\usepackage{pifont}

\long\def\del #1 \enddel { }

\usepackage{amsmath}
\usepackage{graphicx}
\usepackage{hyperref}

\usepackage{color} 

\usepackage{xcolor,colortbl}
\definecolor{Gray}{gray}{0.85}
\definecolor{LightGray}{gray}{0.93}
\definecolor{LightGreen}{rgb}{0.88, 1, 0.88}
\definecolor{LightCyan}{rgb}{0.88,1,1}
\definecolor{LightRed}{rgb}{1, 0.85, 0.85}
\definecolor{LightYellow}{rgb}{1, 1, 0.85}
\definecolor{LightBlue}{rgb}{0.87, 0.94, 1}
\definecolor{white}{gray}{1}

\usepackage{array,mathtools,amssymb,booktabs}
\newcolumntype{G}{>{\columncolor{LightGray}}c}
\newcolumntype{C}{>{$}c<{$}}
\AtBeginDocument{
\heavyrulewidth=.16em
\lightrulewidth=.1em
\cmidrulewidth=.03em
\belowrulesep=.4ex
\belowbottomsep=0pt
\aboverulesep=.4ex
\abovetopsep=0pt
\cmidrulesep=\doublerulesep
\cmidrulekern=.5em
\defaultaddspace=.5em
}

\def\beq{\begin{equation}}
\def\eeq{\end{equation}}

\def\bea{\arraycolsep .1em \begin{eqnarray}}
\def\eea{\end{eqnarray}}
\def\Tr{{\rm Tr}}

\def\eq#1{(\ref{#1})}

\def\s0#1#2{\mbox{\small{$ \frac{#1}{#2} $}}}
\def\0#1#2{\frac{#1}{#2}}

\def\grgl{\:\hbox to -0.2pt{\lower2.5pt\hbox{$\sim$}\hss}{\raise3pt\hbox{$>$}}\:}
\def\klgl{\:\hbox to -0.2pt{\lower2.5pt\hbox{$\sim$}\hss}{\raise3pt\hbox{$<$}}\:}

\makeatletter
    \def\CT@@do@color{%
      \global\let\CT@do@color\relax
            \@tempdima\wd\z@
            \advance\@tempdima\@tempdimb
            \advance\@tempdima\@tempdimc
    \advance\@tempdimb\tabcolsep
    \advance\@tempdimc\tabcolsep
    \advance\@tempdima2\tabcolsep
            \kern-\@tempdimb
            \leaders\vrule
                    \hskip\@tempdima\@plus  1fill
            \kern-\@tempdimc
            \hskip-\wd\z@ \@plus -1fill }

\begin{document}
\title{Advances in  asymptotic safety of $f(R)$ quantum gravity}
\title{Aspects of  asymptotic safety in $f(R)$ quantum gravity}
\title{Aspects of  asymptotic safety for quantum gravity}
\author{Kevin G.~Falls}
\address{\mbox{Scuola Internazionale di Studi Superiori Avanzati (SISSA), via Bonomea 265, 34136 Trieste, Italy.}}
\address{INFN, Sezione di Trieste, Italy}
\author{Daniel F.~Litim}
\author{Jan~Schr\"oder}
\address{\mbox{Department of Physics and Astronomy, University of Sussex, Brighton, BN1 9QH, U.K.}}

\begin{abstract}
We study fixed points of quantum gravity with renormalisation group methods, and a procedure to remove convergence-limiting poles from the flow. The  setup is tested within the $f(R)$ approximation for gravity by solving exact recursive relations up to order $R^{70}$ in the Ricci scalar, combined with resummations and numerical integration. Results include fixed points, scaling exponents, gap in the eigenvalue spectrum, dimensionality of the UV critical surface, fingerprints for weak coupling, and quantum equations of motion. Our findings strengthen the view that ``most of quantum gravity'' is rather weakly coupled. Another novelty are a pair of de Sitter solutions for quantum cosmology, whose occurrence is traced back to the removal of poles. We also address  slight disparities of results in the literature, and give bounds on the number of fundamentally free parameters of quantum gravity.
 \end{abstract}
\maketitle

\tableofcontents

\newpage
\section{\bf Introduction}

Despite of its perturbative non-renormalisability  it continues to be an  open question whether  Einstein gravity  
 exists as a well-behaved and predictive quantum theory of the metric degrees of freedom \cite{Weinberg:1980gg}.
 At low energies, quantum gravity effects can be determined reliably. With increasing energy, however, the  interaction strength grows without bound and an effective theory description breaks down near the Planck scale  \cite{Donoghue:1993eb}.
 The main novelty of the asymptotic safety conjecture for gravity is to observe that quantum effects turn gravitational couplings such as Newton's constant into  running ones
\cite{Weinberg:1980gg,Litim:2006dx,Niedermaier:2006ns,Litim:2011cp}. 
The anti-screening nature of gravity thereby weakens  the gravitational couplings at shortest distances and leads to an interacting fixed point above the Planck scale \cite{Litim:2006dx}. 
A finite number of fundamentally free parameters -- associated with the relevant  interactions in the UV -- ensure predictivity, and  characterise renormalisation group  trajectories which connect the high-energy fixed point  with  classical general relativity at low energies.

 Wilson's renormalisation group \cite{Wilson:1973jj,Polchinski:1983gv,Wetterich:1992yh} has become a powerful tool for tests of the asymptotic safety conjecture \cite{Reuter:1996cp}. 
The methodology is highly versatile  \cite{Litim:1998nf,Berges:2000ew,Pawlowski:2005xe} and allows the derivation of flow equations for running couplings 
at weak and strong coupling alike.
In asymptotically safe theories, and much unlike in asymptotically free ones, the set of relevant, marginal, and irrelevant couplings  is not known from the outset. Still, interacting fixed points can  be searched for  systematically  with the help of  a bootstrap  \cite{Falls:2013bv}. 
In this spirit, and with canonical power counting as the primary guiding principle,
substantial  evidence for asymptotic safety has been accumulated. This covers  Einstein Hilbert approximations and variants thereof
\cite{
Souma:1999at,
Souma:2000vs,
 Reuter:2001ag,
 Lauscher:2001ya,
 Litim:2003vp,
 Bonanno:2004sy,
 Fischer:2006fz,
 Litim:2008tt,   
  Eichhorn:2009ah,
 Manrique:2009uh,
  Eichhorn:2010tb,
  Manrique:2010am,
  Manrique:2011jc,
  Litim:2012vz,
  Donkin:2012ud,
  Christiansen:2012rx,
  Codello:2013fpa,
  Christiansen:2014raa,
  Becker:2014qya,
  Falls:2014zba,
  Falls:2015qga,
  Falls:2015cta,
  Christiansen:2015rva,
  Gies:2015tca,
  Benedetti:2015zsw,
  Biemans:2016rvp,
  Pagani:2016dof,
  Denz:2016qks,
  Falls:2017cze,
  Houthoff:2017oam,
  Knorr:2017fus},
  numerous higher curvature extensions 
  \cite{
Lauscher:2002sq,
  Codello:2006in,
  Codello:2007bd,
  Machado:2007ea,
  Codello:2008vh,
  Niedermaier:2011zz,
 Niedermaier:2009zz,
 Niedermaier:2010zz,
 Benedetti:2009rx,
  Benedetti:2012dx,
  Dietz:2012ic,
  Falls:2013bv, 
  Ohta:2013uca,
 Benedetti:2013jk,
 Dietz:2013sba,
  Falls:2014tra,
Saltas:2014cta,
   Eichhorn:2015bna,
  Ohta:2015efa,
  Ohta:2015fcu,
Demmel:2015oqa,
Falls:2016wsa,
  Falls:2016msz,
  Gies:2016con,
  Christiansen:2016sjn,
  Gonzalez-Martin:2017gza,
  Becker:2017tcx,
  Falls:2017lst,
  deBrito:2018jxt}, 
and  the inclusion of matter 
\cite{
Percacci:2002ie,
Percacci:2003jz,
Narain:2009fy,
Daum:2010bc,
Folkerts:2011jz,
Harst:2011zx,
Eichhorn:2011pc,
Eichhorn:2012va,
Dona:2012am,
Henz:2013oxa,
Dona:2013qba,
Percacci:2015wwa,
Labus:2015ska,
Oda:2015sma,
Meibohm:2015twa,
Dona:2015tnf,
Schroeder:Thesis,
Meibohm:2016mkp,
Eichhorn:2016esv,
Henz:2016aoh,
Eichhorn:2016vvy,
Christiansen:2017gtg,
Eichhorn:2017eht,
Biemans:2017zca,
Christiansen:2017qca,
Eichhorn:2017ylw,
Eichhorn:2017lry,
Christiansen:2017cxa} (see \cite{Eichhorn:2018yfc} for a recent review).
High-order studies  have also revealed  that quantum gravity becomes ``as Gaussian as it gets'',  meaning mostly weakly coupled except for a few dominant interactions \cite{Falls:2013bv,Falls:2014tra,Falls:2017lst}. 
In a similar vein, for general  $4d$ quantum field theories  {\it without}  gravity, 
 strict conditions and  no-go theorems
for asymptotic safety at weak coupling have also been found  \cite{Bond:2016dvk,Bond:2018oco}. Interacting fixed points  in gauge-matter theories \cite{Litim:2014uca,
Bond:2017sem,Buyukbese:2017ehm,Bond:2017tbw,Bond:2017wut,Bond:2017suy,Bond:2017lnq} are invariably  in accord with the bootstrap. Further examples for exact asymptotic safety include large-$N$ fermionic, scalar, Yukawa, and supersymmetric models in $3d$ \cite{
Gies:2009hq,Braun:2010tt,Litim:2011bf,Heilmann:2012yf,
Marchais:2017jqc,
Litim:2018pxe,Jakovac:2013jua,Jakovac:2014lqa,Litim:2016hlb}.

An important  recent stream of works has dealt with simplified models with $f(R)$-type actions for quantum gravity.  A  motivation for this is the
similarity with derivative expansions in critical scalar theories \cite{Benedetti:2012dx}, where  fixed points are under good control \cite{Litim:2010tt}. Thus far studies include high-order polynomial approximations for small curvature,
Pad\'e\ resummations, asymptotic expansions for large  curvature, and fully integrated solutions.
These works  also brought a number of  interesting riddles to light. Most studies agree on the existence of fixed points,
yet some  fail to identify scaling exponents in the first place, while those that do determine them then fail to agree
on the precise number of relevant couplings and the dimension of the UV critical surface. Also, small field and large field  expansions find   vastly different scaling exponents. 
Many works have noted  that gravitational flows  employing  background fields  display technical singularities in the Ricci curvature. The requirement of regularity for physical solutions 
across singular points constrains, and potentially overconstrains, the feasibility of  fixed points. Finally, in some settings the fixed point equations of motion display de Sitter solutions while in others they do not.

In this paper, we have a fresh look at gravitational flows.
Our  primary focus are improved momentum cutoffs, constructed in such a way that technical poles along the flow are absent from the outset \cite{Benedetti:2012dx}.  
In this manner, we find flows with the smallest number of  poles,  and  the smallest number of constraints. We are particularly keen on clarifying whether and how the absence of poles modifies  fixed points, and whether this affects in any noticeable way some of the key features such as the weakly-coupled nature of higher order interactions.  

We  exemplify  our strategy within high-order polynomial expansions of $f(R)$ quantum gravity, doubling-up over previous efforts \cite{Schroeder:Thesis}. The stability and good convergence observed earlier  \cite{Falls:2013bv,Falls:2014tra,Falls:2016wsa} makes this setup an ideal starting point for our purposes. It also allows a detailed comparison with earlier work where technical poles were still present, covering all salient features of the theory including fixed points, radii of convergence, scaling exponents, gap in the eigenvalue spectrum, near-Gaussianity, and de Sitter solutions. For reasons detailed below, we limit ourselves to the regime of small or moderate Ricci curvature, up to the order of the renormalisation group scale. 
In addition, we provide bounds on the fundamentally free parameters of  quantum gravity and offer a perspective on how some of the differences amongst earlier findings can be lifted. Some of our findings have structural similarities with interacting fixed points in non-gravitational theories, and we add a few comments to that effect whenever suitable.

 The rest of the paper is organised as follows. In Sec.~\ref{Setup} we review the derivation of flow equations for $f(R)$ gravity (Sec.\ref{flow}), explain the origin of curvature poles along the flow, detail our removal strategy (Sec.~\ref{removal}), and introduce our  main new modifications 
 (Sec.~\ref{improved}).  In Sec.~\ref{results} we present and discuss our results for asymptotic safety of gravity, including high-order recursive solutions for polynomial fixed points   in the curvature scalar (Sec.~\ref{FP}), details of the underlying bootstrap search strategy (Sec.~\ref{bootstrap}), results for the size of the UV critical surface (Sec.~\ref{UVcs}), bounds on the number of fundamentally free parameters of quantum gravity  (Sec.~\ref{bounds}), evidence for near-Gaussianity of universal scaling exponents (Sec.~\ref{NearGaussian}), some fingerprints for weak coupling  (Sec.~\ref{perturbative}), and  new implications for cosmology (Sec.~\ref{Inflation}).
We disucss our findings and conclude in Sec~\ref{Conclusion}.

\section{\bf Renormalisation group} 
\label{Setup}
In this section we discuss the renormalisation group for gravity and  a new setup leading to improved flow equation, exemplarily given for $f(R)$ models of quantum gravity. 

\subsection{Flows for quantum gravity}\label{flow}
We put forward exact renormalisation group equations \cite{Wetterich:1992yh} adapted for quantum gravity 
following the methods of 
\cite{Reuter:1996cp,Litim:2003vp,Codello:2007bd,Codello:2008vh,Machado:2007ea}. The setup centres around the effective average action $\Gamma_k$ which depends on the infrared cutoff scale $k$.
Our ansatz for the effective average action takes the form of a scale dependent $F_k(R)$ theory in $d=4$ spacetime dimensions,
\beq\label{FR}
\Gamma_{k}  = \int d^4x \sqrt{g} \, F_k(R) 
\eeq
supplemented by classical gauge fixing $S_{\rm gf}$, ghost $S_{\rm gh}$ and auxiliary field terms  $S_{\rm aux}$.
The function $F_k(\bar{R})$ depends on the RG scale $k$ as well as the (dimensionful) scalar curvature $\bar{R}$. The additional terms are assumed to be independent of the scale $k$ taking the form of the corresponding terms in the bare action. We also introduce a wilsonian cutoff by adding a term quadratic in the fields to the action,
\beq\label{Sk}
\Delta S_k[\varphi]=\frac12\int\,\frac{d^dq}{(2\pi)^d}\,\varphi_i(-q)\, \mathcal{R}^{ij}_k(q^2)\,\varphi_j(q)\,.
\eeq
Here $\{\varphi_i\}$ stands for the different fields and $\mathcal{R}_k$ for the matrix of wilsonian momentum cutoffs in field space. It obeys the limits $\mathcal{R}_k(q^2)>0$ for $q^2/k^2\to 0$ and $\mathcal{R}_k(q^2)\to 0$ for $k^2/q^2\to 0$. The cutoff \eq{Sk} 
is removed once $k\to 0$ whereby $\Gamma_k$ falls back onto the full physical effective action. Gauge or diffeomorphism invariance is implemented with the help of the background field method where the spacetime metric
\beq
g_{\mu\nu} = \bar{g}_{\mu\nu} + h_{\mu\nu}
\eeq
is split into a fluctuation $h_{\mu\nu}$ and background $\bar{g}_{\mu\nu}$. The exact functional flow for \eq{FR} is given by 
\beq
\label{FRG}
\partial_t\Gamma_k=\frac12\Tr\frac{1}{\Gamma_k^{(2)}+\mathcal{R}_k}\partial_t \mathcal{R}_k\,.
\eeq
It relates the change with renormalisation group ``time'' $t=\ln k$ of the effective action $\Gamma_k$ to an operator trace over the full propagator multiplied with the scale derivative of the cutoff itself.   The momentum cutoff in field space $\mathcal{R}_k$ ensures that \eq{FRG} is finite both in the UV and in the IR.  By construction \eq{FRG} interpolates between  a microscopic  (classical) theory $(k\to\infty)$
and the full quantum effective action $\Gamma$ $(k\to 0)$.
At weak coupling, iterative solutions of the flow \eq{FRG} generate the conventional perturbative loop expansion \cite{Litim:2001ky,Litim:2002xm}. In the limit where the momentum cutoff $R_k(q^2)$ becomes a momentum-independent mass term the flow \eq{FRG} reduces to the well-known Callan-Symanzik equation \cite{Litim:1998nf}. The right-hand side (RHS) of the flow \eq{FRG} is  local in field and momentum space implying that the change of $\Gamma_k$ at momentum scale $k$ is governed by fluctuations with momenta of the order of $k$. Optimised choices for the regulator term \cite{Litim:2000ci,Litim:2001up} allow for analytic flows and an improved convergence \cite{Litim:2005us}. 
In given approximations, the impact of neglected invariants in the effective action can be probed through the variation of technical parameters such as the RG cutoff scheme  ${\cal R}_k$ \cite{Litim:1996nw,Freire:2000sx,
Litim:2001dt,
Litim:2001fd,
Fischer:2006fz} and through variations of the projection method \cite{Litim:2010tt}.

To compute the flow equation for $\Gamma_k$ in gravity with action \eq{FR} we evaluate the trace appearing on the right side of \eq{FRG} on a space of constant curvature and to identify $\bar{g}_{\mu\nu} = g_{\mu\nu}$ \cite{Falls:2017lst}. To facilitate this, we employ the transverse-traceless decomposition of the metric fluctuation
\cite{York:1973ia} 
\begin{equation}
h_{\mu\nu}=h^T_{\mu\nu}+\bar \nabla_{\mu}\xi_{\nu}+\bar \nabla_{\nu}\xi_{\mu}+\bar \nabla_{\mu}\bar \nabla_{\nu}\sigma-\frac{1}{d}\bar g_{\mu\nu}\bar \nabla^2\sigma+\frac{1}{d}\bar g_{\mu\nu}h\ ,\label{MetricDecomposition}
\end{equation}
and similarly for the ghosts
\begin{equation}
C_{\mu}=C_{\mu}^T+\bar \nabla_{\mu}\eta, \qquad \qquad \bar C_{\mu}=\bar C^T_{\mu}+\bar \nabla_{\mu}\bar\eta. \label{GhostDecomposition}
\end{equation}
 After making these decompositions we also obtain contributions to the operator trace from a host of auxiliary fields resulting from the non-trivial Jacobians in the path integral measure following from \eq{MetricDecomposition} and \eq{GhostDecomposition}.
We then obtain several traces over the metric fluctuations, ghosts and auxiliary fields, and  
the flow equation \eq{FRG} with \eq{FR} takes the form
\begin{widetext}
\bea \label{Operator_trace}
\int d^dx  \sqrt{g} \, \partial_t F_k(R) &=&\frac{1}{2}\textnormal{Tr}_{(2T)}\left[\frac{\partial_t\mathcal{R}_k^{h^Th^T}}{\Gamma^{(2)}_{h^Th^T} + \mathcal{R}_k^{h^Th^T} }\right] +\frac{1}{2}\textnormal{Tr}_{(0)}\left[\frac{\partial_t\mathcal{R}_k^{hh}}{\Gamma^{(2)}_{hh} + \mathcal{R}_k^{hh} }\right]+\frac{1}{2}\textnormal{Tr}^{'}_{(1T)}\left[\frac{\partial_t\mathcal{R}_k^{\xi\xi}}{\Gamma^{(2)}_{\xi\xi} + \mathcal{R}_k^{\xi\xi}}\right]+   \nonumber\\ && 
+  \frac{1}{2}\textnormal{Tr}^{''}_{(0)}\left[\frac{\partial_t\mathcal{R}_k^{\sigma\sigma}}{\Gamma^{(2)}_{\sigma\sigma} + \mathcal{R}_k^{\sigma\sigma} }\right]   + \textnormal{Tr}^{''}_{(0)}\left[\frac{\partial_t\mathcal{R}_k^{\sigma h}}{\Gamma^{(2)}_{\sigma h} + \mathcal{R}_k^{\sigma h} }\right]
 -\textnormal{Tr}_{(0)}^{''}\left[\frac{\partial_t\mathcal{R}_k^{\bar\lambda\lambda}}{\Gamma^{(2)}_{\bar \lambda\lambda}   +   \mathcal{R}^{\bar \lambda \lambda} }\right]
\nonumber \\ &&
+\frac{1}{2}\textnormal{Tr}_{(0)}^{''}\left[\frac{\partial_t\mathcal{R}_k^{\omega\omega}}{\Gamma^{(2)}_{\omega\omega} +   \mathcal{R}^{\omega \omega }   }\right]   -\textnormal{Tr}_{(1T)}^{'}\left[\frac{\partial_t\mathcal{R}_k^{\bar c^Tc^{T}}}{\Gamma^{(2)}_{\bar c^Tc^{T}}   + \mathcal{R}^{\bar c^Tc^{T}}   }\right]+\frac{1}{2}\textnormal{Tr}_{(1T)}^{'}\left[\frac{\partial_t\mathcal{R}_k^{\zeta^T\zeta^{T}}}{\Gamma^{(2)}_{\zeta^T\zeta^{T}}   + \mathcal{R}^{\zeta^T \zeta^T}    }\right]  \nonumber \\ &&
    -\textnormal{Tr}_{(1T)}^{'}\left[\frac{\partial_t\mathcal{R}_k^{\bar C^TC^{T}}}{\Gamma^{(2)}_{\bar C^TC^{T}}   + \mathcal{R}^{\bar C^TC^{T}}   }\right]   -\textnormal{Tr}_{(0)}^{''}\left[\frac{\partial_t\mathcal{R}_k^{\bar\eta\eta}}{\Gamma^{(2)}_{\bar\eta\eta}  + \mathcal{R}^{\bar \eta \eta}   }\right] +   \textnormal{Tr}_{(0)}^{''}\left[\frac{\partial_t\mathcal{R}_k^{\bar ss}}{\Gamma^{(2)}_{\bar ss}    + \mathcal{R}^{\bar s s}   }\right]\label{OperatorTraceFR}
\eea 
\end{widetext}
where all the hessians in the traces are evaluated at $\bar{g}_{\mu\nu} = g_{\mu\nu}$. Here $\bar{\lambda}$, $\lambda$ and $\omega$ are auxiliary fields arising from the Jacobian associated to $\sigma$,  $\bar{c}^T$, $c^T$ and $\zeta$ are auxiliary fields arising from the Jacobian associated to $\xi$, and $\bar{s}$ and $s$ are auxiliary fields arising from the Jacobian associated to $\bar{\eta}$ and $\eta$. 
The prime or the double prime   at certain traces in \eq{OperatorTraceFR} indicate that the zero mode or the zero mode and the negative modes in the eigenspectrum of Laplace-type operators on   the $4$-sphere should be removed (see Sec.~\ref{removal}).

In order to find a final form of the flow equation for $F_k(R)$  one needs to specify the gauge fixing condition and the cutoffs $\mathcal{R}_k$ for the various fields.
We will employ the gauge  fixing condition 
\beq\label{gauge}
F_{\mu} =   \sqrt{2}\kappa\left(\bar \nabla^{\nu}h_{\mu\nu}-\frac{1+\delta}{d}\bar \nabla_{\mu}h\right), 
\eeq
for $\delta =0$, with  $S_{\rm gf}=\frac{1}{2\alpha}\int d^dx\sqrt{\bar g}\,\bar g^{\mu\nu}{F}_{\mu}{F}_{\nu}$. We  also take the Landau limit $\alpha \to 0$ which is a renormalisation group fixed point \cite{Litim:1998qi}. For this choice the gauge fixing action is independent of the trace $h = \bar{g}^{\mu\nu} h_{\mu\nu}$ and the transverse-traceless metric fluctuation $h_{\mu\nu}^T$ depending only on $\xi$ and $\sigma$.
For the two ``physical'' fluctuations $h$ and $h_{\mu\nu}^T$ we use a ``type I'' cutoff function \cite{Codello:2007bd,Codello:2008vh} defined by the following replacement formula
\beq \label{TypeI}
\begin{array}{rcl}
\Gamma^{(2)}_{hh} + \mathcal{R}_k^{hh} &=& \Gamma^{(2)}_{hh}(-\nabla^2 \to -\nabla^2 + R_k(-\nabla^2))\,, \\[2ex]
\Gamma^{(2)}_{h^Th^T} + \mathcal{R}_k^{h^Th^T}& =& \Gamma^{(2)}_{h^Th^T}(-\nabla^2 \to -\nabla^2 + R_k(-\nabla^2))\,, 
\end{array}
\eeq 
where $R_k(Z)$ is the shape function which we choose to be the optimised cutoff \cite{Litim:2000ci,Litim:2001up,Litim:2001fd}
\beq \label{Optimised}
R_k(Z) = (k^2-Z) \theta(k^2 -Z) 
\eeq
and $\theta(k^2 -Z)$  the Heaviside theta function.

For the fluctuations other than those of the physical $h$ and $h_{\mu\nu}^T$ fields we shall choose the regulator to achieve two aims: firstly, to  maximise the cancellation of unphysical modes, and, secondly, to minimize regulator-induced artifacts such as technical  poles that are known to arise  for some regulators. The first aim can be satisfied at the level of operator traces. For our choice of gauge, cancellations between terms in \eq{Operator_trace} will occur provided we choose the regulator coherently for different fields. This can be achieved by  picking a type I cutoff for all fields as in \eq{TypeI}, whereby  equations fall back to those investigated previously \cite{Falls:2013bv,Falls:2014tra}  (see also \cite{Codello:2007bd,Codello:2008vh,Machado:2007ea}),
\beq \label{flowtraces_I}
\begin{array}{rcl}
\displaystyle 
\partial_t \Gamma_k &=& 
\displaystyle
\frac{1}{2}  \Tr \frac{\partial_t \mathcal{R}_{h^Th^T}}{\Gamma^{(2)}_{h^Th^T}+ \mathcal{R}_{h^Th^T}}    
+ \frac{1}{2}  \Tr \frac{\partial_t \mathcal{R}_{hh}}{\Gamma^{(2)}_{hh}+ \mathcal{R}_{hh}} 
\\[4ex]&&\displaystyle 
- \frac{1}{2}  \Tr'' \frac{\partial_t R_k(-\nabla^2)}{- \nabla^2  - \frac{R}{d-1} + R_k(-\nabla^2)} 
- \frac{1}{2}  \Tr' \frac{\partial_t R_k(-\nabla^2)}{-\nabla^2 - \frac{R}{d}+ R_k(-\nabla^2)} \,.
\end{array}
\eeq
Cancellations have occurred between the gauge-fixed fields and the corresponding ghosts. 
However the flow equation \eq{flowtraces_I} leads to technical poles  for  positive real Ricci curvature. For  \eq{Optimised} they are located at
\beq\label{pole}
R = (d-1)\cdot k^2\,\quad {\rm and}\quad R=d\cdot k^2\,.
\eeq
This follows from noticing that   the numerator of the last two terms in \eq{flowtraces_I} with \eq{Optimised} vanishes identically as soon as $\theta =0$. When $\theta =1$, their denominators turn into $ k^2 - R/d$ and $k^2 - R/(d-1)$, respectively, leading to \eq{pole}. 
Evidently, the poles \eq{pole} are of a technical nature in that they arise from technical choices which regularise the ``unphysical'' degrees of freedom. From a practical point of view,  singularities or poles constrain the radius of convergence of polynomial expansions \cite{Falls:2016wsa,Falls:2017lst}. Poles for real positive Ricci curvature may also affect the existence of global fixed point solutions (for all fields). Provided that the number of poles is larger than the number of  boundary conditions that can be tuned to avoid singularities, too many technical poles may rule out the existence of a global fixed point  \cite{Benedetti:2012dx,Dietz:2012ic}. For this reason, we will now turn to a setup in which technical poles are absent by construction. 

\subsection{Removal of curvature poles}\label{removal}

We now explain how technical poles for real  Ricci curvature -- which may arise as a consequence of  choices for the momentum cutoff -- are avoided.
An example for this has been given in \cite{Benedetti:2012dx}, where the technical poles \eq{pole} were removed by a choice of regularisation,  alongside further modifications. 
To avoid regulator-induced pathologies we will adopt a cutoff procedure which modifies the third and fourth term in \eq{flowtraces_I}.  Our modifications are minimal, and applicable even beyond the $f(R)$-type actions discussed here.

Let us first note that in the gauge $\delta =0$ the hessian of gauge fixing action and the hessian of the ghost and auxiliary fields takes the form (up to inconsequential constant factors) 
\beq
\begin{array}{l}
 S^{(2)}_{\rm gf \sigma \sigma} = \frac{1}{\alpha} \Delta_{0b}^2\, \Delta_{0a}\,,\quad
S^{(2)}_{{\rm aux}, \bar{\lambda} \lambda }  = S^{(2)}_{{\rm aux}, \omega \omega }  =  \Delta_{0b} \,\Delta_{0a}
\,,\quad
S^{(2)}_{{\rm gh}, \bar{\eta} \eta }  = \Delta_{0a}\, \Delta_{0b}
\,,\quad
S^{(2)}_{{\rm aux,} \bar{s} s } = \Delta_{0a} 
\\[2ex]
S^{(2)}_{{\rm gf}, \xi \xi} = \frac{1}{\alpha} \Delta_1^2   \,,\quad\quad\quad
S^{(2)}_{{\rm gh}, \bar{C}^T C^T } = S^{(2)}_{\rm aux \bar{c}^T c^T }  = S^{(2)}_{\rm aux \zeta \zeta}  = \Delta_1
\end{array}
\eeq
where we have three types of differential operators,
\beq  \label{S2} 
\Delta_{0a} \ = -\nabla^2\,,\quad
\Delta_{0b} =-\nabla^2 - \frac{\bar{R}}{d-1}\,,\quad
\Delta_1= -\nabla^2 - \frac{\bar{R}}{d} \,.
\eeq
We then pick the regulated two-point function $\Gamma^{(2)}+ \mathcal{R}_k$ such that each of the operators \eq{S2} in the hessians is replaced according to 
\beq
\Gamma^{(2)}+ \mathcal{R}_k  =   S^{(2)}(\Delta_i \to \Delta_i  + R_k(\Delta_i) )
\eeq
for $i = 0a, 0b, 1$. This prescription determines the regulator $ \mathcal{R}_k$ in \eq{FRG}.
For example, the longitudinal gauge-fixed field $\sigma$  is regularised according to
\beq\nonumber
S^{(2)}_{{\rm gf}, \sigma \sigma} + \mathcal{R}_{k, \sigma \sigma} = \big[\Delta_{0b} + R_k(\Delta_{0b})\big]^2 \cdot \big[\Delta_{0a }+ R_k(\Delta_{0a})\big] \,. 
\eeq
It then follows from the structure of the flow equation \eq{FRG} that those traces where the hessians are higher-order differential operators turn into  a sum of traces involving only a single second-order differential operator, plus the regulator term in the denominator. This is illustrated again for the longitudinal gauge-fixed field $\sigma$, where  we find
\bea 
\nonumber
\Tr \frac{\partial_t \mathcal{R}_k}{ S^{(2)}_{{\rm gf}, \sigma \sigma}  +   \mathcal{R}_k} 
&=&  
 \partial_t \Tr  \log \left [S^{(2)}_{{\rm gf}, \,\sigma \sigma} + \mathcal{R}_k \right]
=
  2\Tr \frac{\partial_t R_k(\Delta_{0b})}{\Delta_{0b}+ R(\Delta_{0b})}  
+ 
\Tr \frac{\partial_t R_k(\Delta_{0a})}{\Delta_{0a}+ R(\Delta_{0a})}\,.
\nonumber\eea
Using this property, it then follows that the traces in \eq{FRG} over all fields reduce to four traces
\beq \label{flowtraces}
\partial_t \Gamma_k = \frac{1}{2}  \Tr \frac{\partial_t \mathcal{R}_{h^Th^T}}{\Gamma^{(2)}_{h^Th^T}+ \mathcal{R}_{h^Th^T}}    + \frac{1}{2}  \Tr \frac{\partial_t \mathcal{R}_{hh}}{\Gamma^{(2)}_{hh}+ \mathcal{R}_{hh}} - \frac{1}{2}  \Tr'' \frac{\partial_t R_k(\Delta_{0b})}{\Delta_{0b}+ R_k(\Delta_{0b})} - \frac{1}{2}  \Tr' \frac{\partial_t R_k(\Delta_1)}{\Delta_1+ R_k(\Delta_1)}\,.
\eeq
Much unlike \eq{flowtraces_I} we observe that curvature poles are absent in the third and fourth term as a consequence  of our cutoff choice. 
When evaluating the traces, as in \eq{OperatorTraceFR},  we must remember to remove the zero mode and the negative modes in the spectrum of $\Delta_1$ and $\Delta_{0b}$ which occur on the $4$-sphere; as usual, this is indicated by the prime and the double prime assigned to the final two traces, respectively.
The new flow equation \eq{flowtraces}, and the  qualitative  and quantitative differences with respect to \eq{flowtraces_I}, is the central subject of this work.

 \subsection{Improved flows for quantum gravity}\label{improved}

Next, to evaluate the traces we use the early time heat kernel expansion. 
At this point we switch to dimensionless variables and set $d=4$ writing 
\beq
f(R) \equiv  16 \pi \,F(\bar{R})/k^4
\eeq
 with $R \equiv \bar{R}/k^2$ denoting the Ricci scalar in units of the RG scale. We have also introduced a numerical factor to ensure that the dimensionless Newton coupling $g\equiv G_N(k) k^2$ is related to $f$ by $g=-1/f'(0)$, without any further numerical factors. 
In these conventions, the flow \eq{flowtraces}  takes the form 
\begin{align}
\label{eq:flow_f_d4}
\partial_t f-2\, R\,  f'(R)+4\, f (R)= \,I[f](R) 
\end{align}
where a prime denotes a derivative with respect to $R$. The function $I[f](R)$ on the RHS encodes the effects of quantum fluctuations, and depend on higher derivatives of $f$ up to $f^{(3)}$. The explicit expressions for $I[f](R)$ are identical to those summarised in App.~A of~\cite{Falls:2014tra}, except for the following substitutions
\beq\label{eq:P0V}
\begin{array}{rl}
\displaystyle
{P_c^S}/{D^S} \to P_0^S &
\displaystyle
= \frac{271}{90}\, R^2  -12\, R - 12\\[2ex]
\displaystyle
{P_c^V}/{D^V} \to  P_0^V &
\displaystyle
= \frac{191}{30}\, R^2 - 24\, R -36\,.
\end{array}
\eeq
 The denominators $D_S = R -3$ and $D_V = R - 4$ which occurred  for a  type I cutoff  in \eq{flowtraces_I} are now absent in \eq{flowtraces}. Structurally, the improvement through \eq{eq:P0V} over previous formulations of the flow relates to the removal of unphysical poles \eq{pole}
which occurred since for a  type I cutoff we have.

Interacting fixed points are the non-trivial solutions $f_*(R)$ of the differential equation $\partial_t f(R) =0$ with \eq{eq:flow_f_d4}.  Using the explicit form of the function $I[f]$ 
 which contains  $f'''(R)$ as highest derivative one can write the fixed point equation in normal form to find a non-linear third-order differential equation for the unknown function $f(R)$ at an interacting fixed point,

\bea \label{eq:fixed_point}
 \frac{df''}{dR}&=&
 \nonumber 
  \frac{(3-R)^2 f''+(3-2 R) f'+2 f}{(\frac{181}{1680} R^4+\frac{29}{15} R^3+\frac{91}{10} R^2-54)R}\times\\[1ex]
  &&  \nonumber  
 \times \Bigg[ 48 \pi   \left(2 f-R f'\right)  
 -  \frac{211 R^2-810 R-1080}{90}  \nonumber
\\[1ex]
&&  
\quad\ \  
-\frac{(\frac{311}{756} R^3-\frac{1}{3}R^2-90  R+240) f'
-(\frac{311}{756} R^3-\frac{1}{6}R^2-30 R+60)R f''}{3 f-(R-3) f'} \\[1ex]
&&\nonumber  
\quad\ \ -\frac{(\frac{37}{756} R^3+\frac{29}{15} R^2+18 R+48) f' - 
(\frac{37}{756} R^4
+\frac{29}{10} R^3
+\frac{121}{5} R^2
+12 R
-216) f''}{(3-R)^2 f''+(3-2 R) f'+2 f} \Bigg]\,.
\eea 
In addition to the pole at $R=0$ we observe that the differential equation \eq{eq:fixed_point} displays 
two fixed singularities on the real axis corresponding to the non-trivial 
zeros of the polynomial in the numerator on the RHS of \eq{eq:fixed_point}.
These are located at
\beq \label{poles}
\begin{array}{rl}
R_{-} &= -9.9985\cdots\\[1ex]
R_+ &= \ \ 2.00648\cdots \,.
\end{array}
\eeq
A few comments on \eq{poles} are in order. 

First and foremost, propagator-induced poles at fixed curvature such as  \eq{pole}  
are  now manifestly absent from the flow. The remaining poles \eq{poles} do not originate  from  propagators. Instead, they invariably relate to the physical trace fluctuations of  the $s$ field $(s = h - \nabla^2 \sigma)$ \footnote{The choice of gauge $\delta=0$ and $\alpha=0$ used in this paper entails $\sigma=0$ and $s=h$.},  and can  be seen as a manifestation of the underlying physical degrees of freedom.

Secondly, for a regular fixed point solution to exist for all fields,
 regularity conditions must be obeyed across the fixed singularities. Technical poles \eq{pole} together with those at \eq{poles} may overconstrain  \eq{flowtraces_I}
and prohibit the existence of global solutions \cite{Dietz:2012ic}.
Given that the fixed point equation \eq{eq:fixed_point} is of third order, the presence of three fixed poles would seem to imply a discrete number of non-singular global  solutions. 
The  distribution of poles \eq{poles} along the real axis, with one on either side of the origin,
implies that a one-parameter family of global solutions may exist for either positive or negative Ricci curvature.\footnote{However, variations of the regulators may shift  the location onto one side of the origin
\cite{Demmel:2015oqa}. Accordingly, one expects a discrete number of global solutions for  backgrounds with either positive or negative curvature.} Regardless of their precise location on the real axis it would seem advantageous to have exactly three fixed singularities without being under- nor overconstrained, and  a discrete number of  isolated  global solutions.
Three fixed singularities is exactly what we find once the technical poles are removed from the flow.

Finally, fixed point solutions often have further singularities (poles or cuts) in the complex field plane. The singularity closest to the origin determines radii of convergence for polynomial expansions of the effective action or the quantum equations of motion. These may well be in the complex Ricci curvature plane such as in \cite{Falls:2016wsa}, or at the points \eq{poles} which lie along the real axis \cite{Falls:2017lst}.

Below, we shall not be concerned with global solutions. 
Primarily, we wish to focus on the regime with small  background curvature  where an expansion of the fundamental action in field monomials 
is viable thanks to  abundant evidence for the applicability of canonical power counting 
\cite{Weinberg:1980gg,Falls:2013bv,Falls:2014tra,Falls:2017lst}. Also, we evaluate the operator traces using the heat kernel, which itself is reliable for sufficiently small background Ricci curvature. 
Finally, we note that little is known about  which curvature invariants control  the large-field asymptotics at a gravitational fixed point.  Requiring fixed point solutions for {\it truncated} effective actions to exist globally  presupposes that the large-field asymptotics is also controlled by the very same {\it truncated} effective action as the small-field region. Since $f(R)$ approximations are not closed, this tacit   assumption cannot be taken for granted.

 \section{\bf Results}
 \label{results}
 Next, we give an overview of results, covering the fixed point from recursive relations, a bootstrap test for asymptotic safety, universal scaling exponents, the dimensionality of the UV critical surface,  the approach to near-Gaussian behaviour, and implications for cosmology. 
\subsection{Fixed points} \label{FP}

We begin by explaining how viable interacting fixed point solutions with \eqref{eq:fixed_point}  are identified 
within a polynomial approximation of the $f(R)$-type action  around vanishing Ricci curvature. To this end, we approximate the action by curvature invariants with increasing canonical mass dimension,
\beq\label{eq:f_expansion}
f(R) =\sum \limits_{n=0}^{N-1}\, \lambda_n\, R^n
\eeq
where the  gravitational couplings are defined as  $\lambda_n = f^{(n)}(0)/n!$, and $N$ denotes the  order of the polynomial approximation.
For sufficiently large $N$, we expect the series to converge within a finite radius $|R| < R_c$.
The ansatz \eq{eq:f_expansion} tacitly assumes that canonical power counting continues to be a good approximation scheme at interacting fixed points, which is confirmed {\it a posteriori} (see below).

At the heart of our  calculations are exact recursion relations for the gravitational  couplings in \eq{eq:f_expansion}. Differentiating the fixed point condition \eqref{eq:fixed_point} and employing the expansion \eqref{eq:f_expansion}, we obtain explicit expressions which determine higher order couplings 
in terms of lower order ones, 
$\lambda_{n+2} = \lambda_{n+2}(\lambda_0, \lambda_1, \dots, \lambda_{n+1})
$. 
Solving these relations recursively yields algebraic expressions for all couplings $\lambda_n$ (for all $n\ge 2)$ in terms of the vacuum energy $\lambda_0$ and (minus) the inverse dimensionless Newton coupling $\lambda_1$, leading to
$\lambda_{n} = \lambda_{n}(\lambda_0,\lambda_1)$.
Closed analytical solutions for the recursive relations are presently not at hand. Still, they can be  solved algebraically, order by order, which we do with the help of a dedicated {\tt C++} code \cite{Schroeder:Thesis}. The resulting expressions have the form of ratios of  polynomials $P_n$ and $Q_n$ in the parameters $\lambda_0$ and $\lambda_1$ with integer coefficients,
\beq\label{lambdan}
\lambda_n=\frac{P_n(\lambda_0,\lambda_1)}{Q_n(\lambda_0,\lambda_1)}\,.
\eeq
We have computed the  polynomials $P_n$ and $Q_n$ up to order $n=70$.  Their size  grows rapidly with $n$: the number of non-trivial terms 
contained in each of them scales approximately as 
$\propto n^{2.7}$ 
with a proportionality constant of the order of a few,  slightly larger for $P_n$ than for $Q_n$. At the highest order, we find $\gtrsim 10^5$ non-trivial terms in either of them. Also, the maximal power of $\lambda_0$ ($\lambda_1$) contained in $P_n$ grows roughly linearly with $n$, and reaches approximately 600 (400) at  $n=70$.

\begin{figure}
\begin{center}
 \vskip-.5cm
    \includegraphics[width=0.5\textwidth]{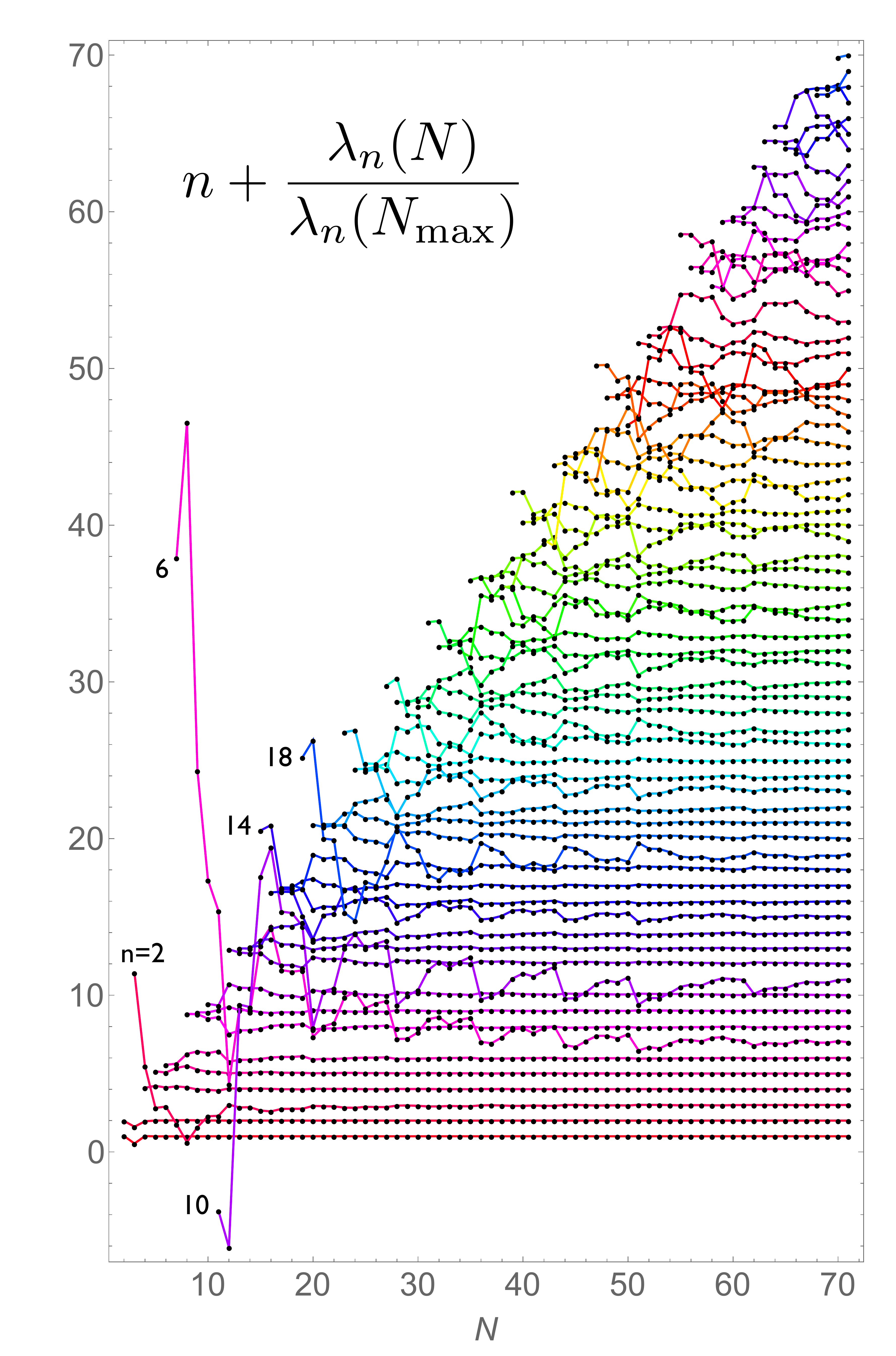}
    \end{center}
\vskip-.5cm
\caption{Convergence of couplings. The  couplings $\lambda_n(N)$ at each approximation order $N$ are shown in units  of their value at the highest order $\lambda_n(N_{max})$, with $N_{\text{max}}=71$.
For better visibility, couplings are shifted by an additional offset $n$. 
 The convergence pattern of all couplings exhibit an underlying eightfold periodicity. 
 Some couplings converge more slowly than others, e.g.~$\lambda_2$, $\lambda_6$ and $\lambda_{10}$ (see main text).
}\label{fig:couplings}
\end{figure}

It remains to determine the  free parameter $\lambda_0$ and $\lambda_1$, for which  we adopt the numerical search strategy detailed in \cite{Falls:2014tra}.  Specifically, we determine the free parameters at polynomial approximation order $N$ by requiring that the next two higher fixed point couplings vanish,
 \begin{equation}
 \begin{array}{rcl}
 0&=&\lambda_N (\lambda_0,\lambda_1)\,,\\[1ex]
 0 &=& \lambda_{N+1}(\lambda_0,\lambda_1)\,. 
 \end{array}
 \label{eq:boundary_condition}
 \end{equation}
 This closes the approximation and provides us with two equations for two unknowns.\footnote{See  \cite{Falls:2014tra} for a discussion of refined boundary conditions.}  Notice that neither the existence nor uniqueness of solutions is guaranteed from the outset.  Still, our search strategy starts off with the Einstein-Hilbert approximation $N=2$, which does offer a unique fixed point candidate. We then increase $N\to N+1$ and search for solutions of the new boundary condition \eq{eq:boundary_condition} which are nearby to the previously found solution. If several such solutions are found, we  pick the one whose scaling exponents are nearby those of the solution found at the  previous approximation order.  In principle, given the algebraic nature of the equations, many solutions may exist. In practice, we only find a small number of solutions at each and every order. With increasing $N$, the physical solution materialises as an accumulation point of nearby spurious solutions \cite{Falls:2014tra}, in a pattern similar to what has been observed previously in critical scalar theories \cite{Litim:2016hlb}. Unphysical solutions appear and disappear with varying $N$, and distinguish themselves through strong variations in the scaling exponents with no consistent convergence pattern from order to order.   More details on our  search routine and  improved  boundary conditions to \eq{eq:boundary_condition} are discussed in \cite{Falls:2014tra}.
Here, we performed this analysis from $N=2$ up  to $N=71$ to reach significantly beyond the maximal order of $N=35$ achieved previously  \cite{Falls:2013bv,Falls:2014tra,Falls:2013bma,Nikolakopoulos:Thesis}.

 Our results for the couplings are displayed in Tab.~\ref{tCouplings}, and Fig.~\ref{fig:couplings}. We find a viable fixed point with a positive Newton's coupling $g=-1/\lambda_1$ and a positive cosmological constant $\lambda=-\lambda_0/(2\lambda_1)$ at each and every approximation order, with Tab.~\ref{tCouplings} showing the first 32 couplings at the highest order $N=71$. 
 All gravitational couplings are of order unity, except for a few couplings $\lambda_{2+4i}$ which tend to come out smaller by one or two orders of magnitude and whose order-by-order convergence is slower. Newton's coupling comes out of order unity, indicating that the fixed point is strongly coupled with sizeable gravitational anomalous dimensions.
Fig.~\ref{fig:couplings} shows the order-by-order convergence of polynomial couplings $\lambda_n$ for $n=0$ up to $n=70$ (bottom to top). For better display, we have normalised couplings in units of their asymptotic values at approximation order $N=71$ (see  Tab.~\ref{tCouplings}).  We observe that couplings converge towards stable values, albeit quite slowly. Roughly every fourth coupling $\lambda_{2+4i}$ displays stronger fluctuations, starting with $\lambda_2$. This can be traced back to the periodicity pattern and the fact that $\lambda_2$ is classically marginal. The absence of a tree-level contribution from the canonical mass dimension  implies that its fixed point must arise purely on the quantum level. Then, cancellations are more sensitive to accurate fixed point values, leading to stronger fluctuations and   slower convergence. 
For the first three couplings, we find that the accuracy in the fixed point  improves by a decimal point if, approximately, $N\to N+40$. 
A similarily slow rate of convergence has been observed in previous $f(R)$ studies, where this has been traced back to a convergence-limiting pole in the complexified field plane \cite{Falls:2013bv,Falls:2014tra,Falls:2016wsa}.

\begin{table}[t]
\begin{tabular}{GrGrGrGr}
\toprule
$\ \lambda _0\ $&$0.2463$ & $\lambda _8$&$0.2648$ & $\lambda _{16}$&$0.4637$ & $\lambda _{24}$&$1.2000$ \\
$\lambda_1$& $-1.0774$ & $\lambda_9$&$0.2017$ & $\lambda _{17}$&$0.3114$ &$\lambda_{25}$&$0.6528$ \\
$\lambda_2$&$0.0085$ & $\lambda_{10}$&$0.0099$ & $\lambda_{18}$&$-0.0570$ & $\lambda_{26}$&$-0.3836$ \\
$\lambda_3$&$-0.4795$ & $\lambda_{11}$&$-0.2364$ & $\lambda_{19}$&$-0.5134$ & $\lambda_{27}$&$-1.4760$ \\
$\lambda_4$&$-0.3755$ & $\lambda_{12}$&$-0.3643$ & $\lambda_{20}$&$-0.7805$ & $\lambda_{28}$&$-2.0443$ \\
$\lambda_5$&$-0.2271$ & $\lambda_{13}$&$-0.2491$ & $\lambda_{21}$&$-0.5397$ & $\lambda_{29}$&$-1.3342$ \\
$\lambda_6$&$0.0059$ & $\lambda_{14}$&$0.0567$ & $\lambda_{22}$&$0.2105$ & $\lambda_{30}$&$0.8319$ \\
$\lambda_7$&$0.1957$ & $\lambda_{15}$&$0.3575$ & $\lambda_{23}$&$0.9892$ & $\lambda_{31}$&$3.1400$ 
\\
\bottomrule
  \end{tabular}
\caption{Fixed point values for the first 32 polynomial couplings. 
All of them are of order unity except $\lambda_2$, $\lambda_6$, $\lambda_{10}$, $\lambda_{14}$, and $\lambda_{18}$ which come out smaller by one to two orders of magnitude, also showing slower convergence (Fig.~\ref{fig:couplings}). We  observe an approximate eight-fold periodicity pattern 
 in the signs of couplings (see main text).} \label{tCouplings}
\end{table}

We also observe an approximate eight-fold periodicity pattern in the sign of couplings, related to a convergence-limiting pole (or cut) of the solution  $f_*(R)$  in the plane of complexified Ricci curvature. The pole is approximately located at an angle of  $\approx \pi/4$ from the origin, and at a distance $R_c$ set by the  radius of convergence \cite{Falls:2014tra,Falls:2016wsa}.  An estimate for the radius of convergence is obtained by evaluating
$R_c = \left({\lambda_n}/{\lambda_{n+m}}\right)^{1/m}$
for each approximation $N$, and in the limit of large $n$ with $m$ fixed.  Taking  $m$ to be an integer multiple of eight to reflect the underlying periodicity pattern and averaging over the accessible values for $n$  we find 
\begin{align} 
R_c\approx 0.878 \label{eq:radius_of_convergence_pure_gravity}
\end{align}
as the mean value over all possible radii. To obtain \eq{eq:radius_of_convergence_pure_gravity}, and at each approximation order $N$, we excluded the five lowest couplings (because the periodicity pattern sets in after that)  and the 10 highest couplings (as these have not yet converged properly). The radius compares well with the result of \cite{Falls:2016wsa}, where a different regularisation was used. 
Moreover, we also confirm another qualitative result of \cite{Falls:2016wsa}, which is that gravity remains an attractive interaction for all background Ricci curvature $(f'(R)<0)$ within $|R|<R_+$.

\subsection{Bootstrap}
\label{bootstrap}
With a stable fixed point candidate at hand, we are now in a position to check whether it is also a viable candidate for an asymptotically safe version of quantum gravity. Since the fixed point is (strongly) interacting, the irrelevancy of higher-dimensional curvature invariants  cannot be taken for granted. To ensure that terms such as $\int\sqrt{g}R^{n}$ with $n>70$ (neglected in the present study) are not spoiling asymptotic safety, we must check that the underlying approximation scheme \eq{eq:f_expansion} 
is viable  \cite{Falls:2013bv}. The tacit hypothesis underneath  \eq{eq:f_expansion}  is that  
\begin{equation}\label{H1}
\bullet 
\begin{array}{rl}
&{\rm 
the\ relevancy\ of\ invariants\ at\ an\ interacting\ fixed\ point\ continues}\\ &{\rm to\ be\ governed\ by\ the\ invariant's\ canonical\ mass\ dimension}\,.
\end{array}
\end{equation}
If so, it is perfectly acceptable to truncate the series \eq{eq:f_expansion}  at some highest order $N_{\rm max}$, such that all relevant operators are already contained within the approximation with no other relevant operartors arising beyond $N_{\rm max}$. The hypothesis  is trivially true for non-interacting fixed points such as in asymptotic freedom, and  for any weakly interacting fixed point where anomalous dimensions are under perturbative control.  It also holds true for all fixed points observed in nature, including strongly interacting ones. 

\begin{figure*}[t]
    \begin{center}
    \includegraphics[width=.7\textwidth, angle=0]{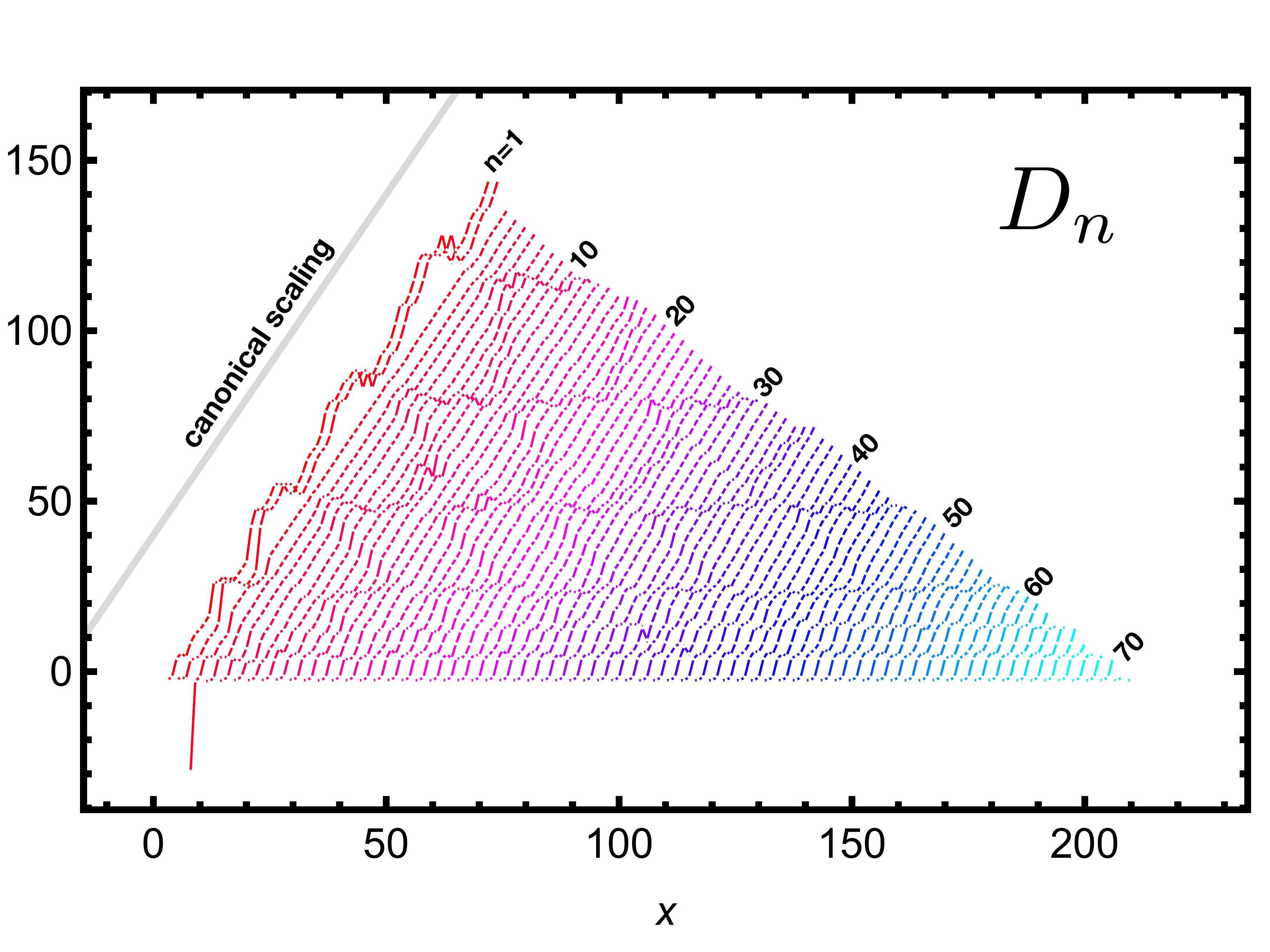}
 \end{center}
\vskip-.5cm
\caption{Bootstrap test for asymptotic safety and  canonical power counting.
From left to right, each line $D_n$ shows the $n^{\rm th}$ largest eigenvalues from all approximation orders $(N\ge n)$, connected by a line. At fixed $n$, and with increasing approximation order (increasing $x=N+2n$)
we observe that all curves $D_n$ consistently grow with an average slope of $
D'_n(x)\approx 2$. This  demonstrates that invariants with increasingly large canonical mass dimension correspond  to increasingly irrelevant operators at the UV fixed point, confirming the  hypothesis \eq{H1}.}
\label{fig:eigenvalues_bootstrap}
\end{figure*}

The virtue of the hypothesis \eq{H1}, for our purposes, is that it can be verified as soon as fixed points and scaling exponents are known. The procedure has   the form of a bootstrap.  
To that end, we must compute the set of universal scaling exponents of the theory at each and every order in the approximation.
Universal scaling exponents are obtained as eigenvalues of the  stability matrix $M_{ij} = {\partial \beta_i}/{\partial \lambda_j}|_*$.
At approximation order $N$, the set of $N$ eigenvalues is given by
\beq\label{theta}
\{ \vartheta_i(N) \big| \ i=0,\cdots,N-1,\ {\rm with}\ {\rm Re}\,  \vartheta_i\le {\rm Re} \,\vartheta_{i+1} \}\,,
\eeq
with eigenvalues sorted according to magnitude. The number of negative eigenvalues determines the size of the UV critical surface \cite{Weinberg:1980gg}, with each negative eigenvalue corresponding to a fundamentally free parameter of the theory. It is vital for the asymptotic safety conjecture that the number of negative eigenvalues remains finite to ensure predictivity. To check whether the hypothesis is realised in the data, we must show that  the step from $N\to N+1$ (corresponding to the inclusion of a new invariant $\int\sqrt{g}R^{N}$) leads to the appearance of a new, largest eigenvalue in the spectrum \eq{theta}, larger than those encountered at lower orders in the approximation \cite{Falls:2013bv,Falls:2014tra}.. 

In Fig.~\ref{fig:eigenvalues_bootstrap}  we summarize the evidence in support of \eq{H1}. The order-by-order variation of eigenvalues is displayed as follows: Each line $D_n(x)$ shows the $n^{\rm th}$ largest eigenvalue in the spectra \eq{theta} (real part if complex),  connected by a line, with  $x(N)=N+2n$ increasing with approximation order $N$.  For example, the line $D_1$, the first line to the left, picks  the largest eigenvalue $\max_i \vartheta_i(N) $ for each and every approximation order, and then connects them with increasing  $N=2,3,4 \cdots,70$ (from bottom left to top right).  The working hypothesis \eq{H1} is verified  provided the lines $D_n$ grow consistently with increasing $x$. This is fully confirmed in Fig.~\ref{fig:eigenvalues_bootstrap} for all lines $D_n$. Notice that, in places, the curves shift sideways, which is understood due to the first occurence of complex conjugate pairs of exponents  \cite{Falls:2014tra}. Overall, the average slope of growth is approximately given by the canonical mass dimension of the Ricci scalar, as expected from canonical power counting. In Fig.~\ref{fig:eigenvalues_bootstrap}, for comparison, the slope corresponding to canonical scaling is indicated by a gray line. We also note that suitably normalised eigenvectors converge well and show the expected pattern with increasing approximation order. Most importantly, there are no indications for the growth to slow down nor to turn over, and we conclude that canonical power counting is applicable in our model for quantum gravity. 

\begin{figure*}[t]
  \begin{center}
  \includegraphics[width=.7\textwidth, angle=0]{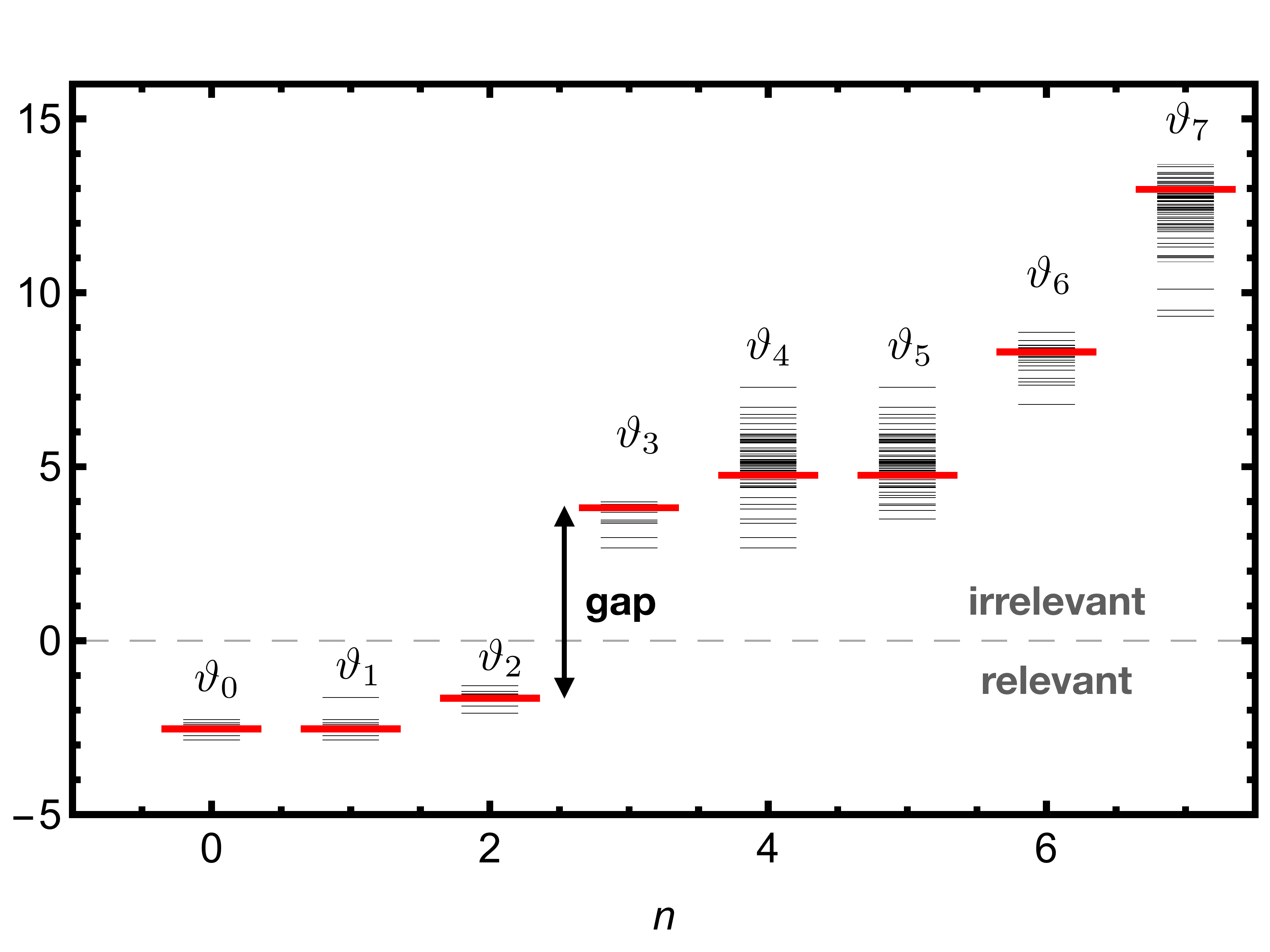}
 \end{center}
\vskip-.5cm
\caption{Dimensionality of UV critical surface. Shown are the smallest eight universal eigenvalues $\vartheta_n(N)$ with $n=0,\cdots,7$ for the 70 different approximation orders (thin black lines), and their values at the highest order (thick red line). All approximation orders invariably display a three-dimensional UV critical surface related to the three negative eigenvalues $\{\vartheta_0,\vartheta_1, \vartheta_2\}$. The conclusion is further supported by the sizeable gap between the smallest negative and the smallest positive eigenvalue.}
\label{fig:UVsurface}
\end{figure*}

\subsection{UV critical surface}
\label{UVcs}
We are now in a position to determine the size of the UV critical surface, determined by the number of negative eigenvalues in the spectrum \eq{theta}. In Fig.~\ref{fig:UVsurface}, we show the first eight eigenvalues for various approximation orders.  Thin black lines indicate the approximately 70 data points, and wider red lines show the values at the highest order.
The first three eigenvalues $\vartheta_0$, $\vartheta_1$, and $\vartheta_2$ come out negative at each and every order in the approximation. Their variation with $N$ is weak. Two of these constitute a complex-conjugate pair, which is in qualitative agreement with earlier findings. 
Introducing $\vartheta_0=-\theta'_0+i\theta''_0$,  
$\vartheta_1=-\theta'_0-i\theta''_0$, and $\vartheta_{2,3}=-\theta_{2,3}$, we find
\beq\label{values}
\begin{array}{rcl}
\langle \theta'_0 \rangle &=& \ \ 2.529\pm 0.018\\[.5ex]
\langle \theta''_0\rangle &=& \ \ 2.196\pm 0.014\\[.5ex]
\langle \theta_2\rangle &=&\ \ 1.657\pm 0.011\\[.5ex]
\langle \theta_3\rangle &=&-3.822\pm 0.015
\end{array}
\eeq
after averaging over the three highest cycles of eight approximations ($i.e.$ the 24 data sets from $N=48$ to $N=71$). The smallness of the error estimate, corresponding to a standard devitation, shows that \eq{values} holds true for essentially all approximation orders individually. From $\vartheta_3$ onwards, all eigenvalues are positive (or have a positive real part, if complex). We conclude that the UV critical surface is three-dimensional.

\begin{figure*}[t]
    \begin{center}
   \includegraphics[width=.7\textwidth, angle=0]{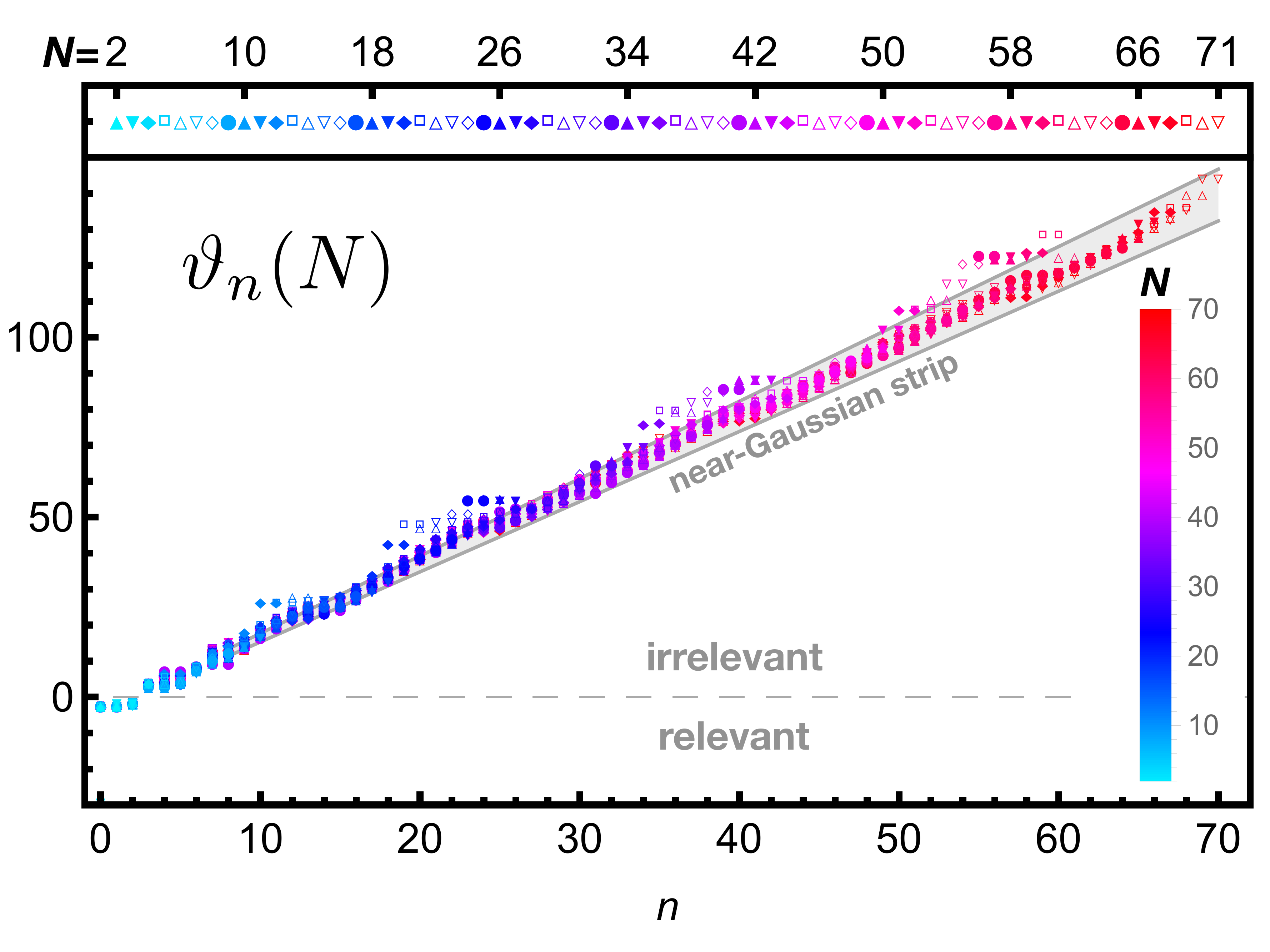}
 \end{center}
\vskip-.5cm
\caption{Universal scaling exponents. Shown are the scaling exponents $\vartheta_n$ for all approximation orders $N$. The inset and the top legend explain the colour-coding of approximation orders $N$. The gray-shaded area indicates that most scaling exponents, with increasing $n$, take values narrowly above, yet increasingly close to classical values within the  near-Gaussian strip.}
\label{fig:UV}
\end{figure*}

Another quantity of particular interest is the gap in the eigenvalue spectrum, $\Delta_{\rm UV}$, which measures the ``distance'' between the smallest relevant and the smallest irrelevant eigenvalues. Classically, the gap reads $\Delta=2$. Quantum-mechanically, we observe from  Fig.~\ref{fig:UVsurface} that $\Delta_{\rm UV}=\vartheta_3-\vartheta_2$. Quantitatively, we find
\beq\label{gap}
\Delta_{\rm UV}
\approx 5.480\pm 0.011\,.
\eeq
Hence, fluctuations have widened the gap. The result  \eq{gap} compare well with findings in \cite{Falls:2014tra}, where $\Delta_{\rm UV}\approx 5.58$ was found. Including Ricci tensor fluctuations, the gap  widens further, $\Delta_{\rm UV}\approx 5.98$. Overall, we conclude from \eq{values} that the leading four scaling exponents deviate noticeably from classical values $\{4,2,0,-2\}$ owing to strong fluctuastions in the deep UV. We also conclude that the modified regularisation of modes introduced here does not lead to qualitative modifications in the results 
\cite{Falls:2013bv,Falls:2014tra,Falls:2013bma,Nikolakopoulos:Thesis}.   

In Fig.~\ref{fig:UV}, we extend the study of Fig.~\ref{fig:UVsurface} to all eigenvalues $\vartheta_n$ for all approximation orders $N$. The inset on the right side  explains the colour-coding of approximation orders $N$, while the top legend indicates the use of symbols. The dashed horizontal line shows the boundary between relevant and irrelevant eigenvalues. For fixed $n$, we note that the variation with $N$ is moderate. 
Also, results for the lower order approximations are plotted on top of those for the higher-order approximations. The near-invisibility of red-coloured data points at low $n$ (Fig.~\ref{fig:UV}) nicely  documents that the low-order approximations only differ very mildly from the more accurate values at higher orders.
Traces of an underlying eightfold periodicity pattern continue to be visible in Fig.~\ref{fig:UV}. 

We note that canonical power counting continues to be a ``good'' ordering principle. Moreover, we also observe that with increasing $n$, most scaling exponents take near-Gaussian  values within or close to the narrow gray-shaded strip  around Gaussian  exponents,
\beq
\vartheta_{G,n} = 2\,n -4 \,. \label{eq:Gaussian_spectrum}
\eeq 
Most of the exponents in the near-Gaussian strip in Fig.~\ref{fig:UV} are slightly above Gaussian values \eq{eq:Gaussian_spectrum}. This behaviour of curvature invariants  with large canonical mass dimension has first been observed in \cite{Falls:2013bv}. We emphasize that a near-Gaussian asymptotics is not required for asymptotic safety. It would be sufficient if scaling exponents $\vartheta_n$ remain positive for sufficiently large $n$. 

The results in Fig.~\ref{fig:UV}  clearly document that the scaling exponents $\{\vartheta_n\}$  at an interacting UV fixed point only deviate noticeably from Gaussian exponents $\{\vartheta_{G,n}\}$  for a rather small set of low-dimensional curvature invariants.  In non-gravitational quantum field theories, Gaussian or near-Gaussian exponents are known to arise either at free or weakly interacting fixed points  in large-$N$  limits, e.g.~\cite{Litim:2014uca,
Bond:2017sem,Buyukbese:2017ehm,Bond:2017tbw,Bond:2017wut,Bond:2017suy,Bond:2017lnq}  \cite{Litim:2011bf,Heilmann:2012yf,Litim:2016hlb,Marchais:2017jqc}. For interacting fixed points, the appearance of exactly classical exponents  is then traced back to the existence of an exactly marginal coupling. In this light, our results
suggest that a ``nearly marginal coupling'' or    ``small parameters'' are hidden within this formulation of quantum gravity. We return to this point in Sec.~\ref{perturbative}.

\subsection{Fundamentally free parameters}\label{bounds}

Let us  discuss our results from the viewpoint of the full quantum theory of gravity beyond the  approximations adopted here. Of particular interest is the set  of fundamentally free parameters of the theory, corresponding to the number  $N_{\rm free}$ of relevant and marginally relevant operators at the UV fixed point \cite{Weinberg:1980gg,Falls:2013bv,Benedetti:2013jk}. To obtain a bound on these, we exploit the applicability of canonical power counting  \cite{Weinberg:1980gg,Falls:2013bv,Falls:2014tra,Falls:2017lst} and
functional RG results from different projection methods and approximation orders. We also take notice of the fact that variations of the RG scheme modify the projection of the full theory onto a given operator basis \cite{Litim:2010tt}. This information may then be used to probe the impact of omitted operators \cite{Litim:1996nw,Freire:2000sx,Litim:2000ci,Litim:2001dt,Litim:2001up,Litim:2010tt}. 

We begin by recalling that the cosmological constant and the Ricci scalar, $i.e.$~the invariants with the smallest canonical mass dimensions,  have robustly been established as relevant interactions irrespective of the underlying RG scheme and the projection method 
\cite{Souma:1999at,
Souma:2000vs,
 Reuter:2001ag,
 Lauscher:2001ya,
 Litim:2003vp,
 Bonanno:2004sy,
 Fischer:2006fz,
Litim:2008tt,   
Eichhorn:2009ah,
 Manrique:2009uh,
  Eichhorn:2010tb,
  Manrique:2010am,
  Manrique:2011jc,
  Litim:2012vz,
  Donkin:2012ud,
  Christiansen:2012rx,
  Codello:2013fpa,
  Christiansen:2014raa,
  Becker:2014qya,
  Falls:2014zba,
  Falls:2015qga,
  Falls:2015cta,
  Christiansen:2015rva,
  Gies:2015tca,
  Benedetti:2015zsw,
  Biemans:2016rvp,
  Pagani:2016dof,
  Denz:2016qks,
  Falls:2017cze,
  Houthoff:2017oam,
  Knorr:2017fus}.
This leads to a first  estimate for a lower bound 
 \beq\label{free2}
  N_{\rm free}\ge 2\,.
 \eeq
Equally important, this result has proven qualitatively and quantitatively robust under the substantial inclusion of higher-dimensional interactions \cite{Falls:2013bv,Falls:2014tra,Falls:2017lst}. We  conjecture that either of these remain relevant couplings in the full theory \cite{Litim:2008tt}. 
 
 Next, following on with the canonically marginal interactions, our results in Figs.~\ref{fig:UVsurface},~\ref{fig:UV} -- based on the projection of fourth-order interactions  onto a $R^2$ term --  state that at least one linear combination of these should become relevant in the full theory. Together with the relevancy of the canonically relevant and marginal couplings, this result implies  the tighter lower bound
 \beq\label{free}
  N_{\rm free}\ge 3
 \eeq
on the number of fundamentally free parameters of quantum gravity. This result is consistent with \cite{Falls:2017lst} where the projection, instead, has been made onto a $R_{\mu\nu}R^{\mu\nu}$ term. 
  In $f(R)$ studies, it has  been observed that   projected $R^2$ interactions may come out irrelevant in particular schemes which utilise the exponential parameterisation of metric fluctuations (or other such non-linear parameterisations) 
\cite{Ohta:2015fcu,deBrito:2018jxt},  or change from relevant to marginal to irrelevant through a continuous variation of the RG scheme 
 \cite{deBrito:2018jxt}, suggesting that $N_{\rm free}$ might stay as low as \eq{free2}.
Still, the eigenvalue spectra may still change under the inclusion of higher order interactions \cite{Falls:2013bv,Falls:2014tra,Falls:2017lst}.
 Moreover, even if they do not, the  findings \cite{Ohta:2015fcu,deBrito:2018jxt} may still be in accord   with  the stronger lower bound \eq{free} since the relevant direction is likely to be a linear combination of the canonically marginal interactions and not a pure $R^2$ term. Hence, modifications of the RG scheme or the projection may push the third relevant interaction  away from the $R^2$ operator and  no longer show up in the spectrum.
 Additionally, results  indicate that at least one linear combination of the fourth order couplings  is irrelevant at a fully interacting fixed point. Incidentally, either of these findings are consistent with the model of  \cite{Benedetti:2009rx}, where it is found with the help of Einstein spaces that  only one of the two linear combinations dominated by $R^2$ and $R_{\mu\nu}R^{\mu\nu}$ interactions become relevant at an interacting fixed point. In this light, the slight disparity amongst some of the earlier findings \cite{Lauscher:2002sq,Codello:2007bd,Machado:2007ea,Falls:2013bv,Falls:2014tra,Ohta:2015fcu,deBrito:2018jxt} is  associated to differences in the projection of the RG flow onto the full theory, rather than to differences in the underlying physics.\footnote{The fate of fourth-order interactions  at  fixed points can be settled in the single-metric approximation by including all of the curvature-squared terms and using generic backgrounds for the evaluation of the operator trace.}

 Finally, it has also been observed that the fourth order interactions may become asymptotically free at the Gaussian fixed point as indicated in perturbation theory, while Newton's coupling and the cosmological constant take an interacting fixed point  \cite{Codello:2008vh,
  Niedermaier:2011zz,
 Niedermaier:2009zz,
 Niedermaier:2010zz}. If so, the lower bound tightens further and reads \cite{Niedermaier:2011zz,
 Niedermaier:2009zz,
 Niedermaier:2010zz}
 \beq\label{free4}
  N_{\rm free}\ge 4\,.
 \eeq
 As such, the partially interacting fixed point is different from the one observed in this work.\footnote{Our study, which finds a finite fixed point for the $R^2$ coupling $\lambda_2$,   is not directly sensitive to the scenario  of a partially interacting  fixed point which requires $1/\lambda_2\to 0$ instead.} It remains to establish that the partially interacting fixed point   \cite{Niedermaier:2011zz,
 Niedermaier:2009zz,
 Niedermaier:2010zz}  persists under the inclusion of higher order interactions \cite{Falls:2013bv,Falls:2014tra,Falls:2017lst}.
 
 Continuing with sixth-order interactions such as $R^3$ \cite{Codello:2007bd,Machado:2007ea,Codello:2008vh,Falls:2013bv,Falls:2014tra},  the Goroff-Sagnotti term 
 $C_{\mu\nu}{}^{\rho\sigma} 
 C_{\rho\sigma}{}^{\tau\lambda}
 C_{\tau\lambda}{}^{\mu\nu}$ \cite{Gies:2016con}, Ricci tensor invariants such as $R \cdot R_{\mu\nu}R^{\mu\nu}$ \cite{Falls:2017lst}, or even higher-order terms such as $R^{2+n}$ \cite{Falls:2013bv,Falls:2014tra},  $ (R_{\mu\nu}R^{\mu\nu})^n$ or $R\cdot (R_{\mu\nu}R^{\mu\nu})^{n}$ \cite{Falls:2017lst},    all for $n\ge 2$, we observe that these invariably come out as irrelevant, including here (Figs.~\ref{fig:UVsurface},~\ref{fig:UV}). In other words, higher-order interactions do not introduce new fundamentally free parameters into the theory, in accord with \eq{free}.  Formal arguments for the existence of an upper bound on $N_{\rm free}$ in the $f(R)$ approximations  \cite{Benedetti:2013jk} are in accord with our results and \cite{Falls:2013bv,Falls:2014tra}. Based on our findings, and together with \cite{Falls:2013bv,Falls:2014tra,Falls:2017lst}, this would seem to indicate  that at most all {classically}  relevant and  marginal couplings become relevant couplings {quantum  mechanically}, though much more work will be required to identify an upper bound on $N_{\rm free}$.
  Hence, provided that the full theory displays an interacting fixed point, our results  offer the lower bound \eq{free} and  indications for an upper bound on the dimensionality of the UV critical surface.\footnote{In uni-modular models of quantum gravity where  diffeomorphisms are restricted to volume-preserving ones, the role of the cosmological constant is modified \cite{Padilla:2014yea} and the bound \eq{free} may be reduced by one \cite{Saltas:2014cta,Eichhorn:2015bna}.}

\subsection{Near-Gaussianity}
\label{NearGaussian}

\begin{figure*}[t]
   \begin{center}
    \includegraphics[width=.93\textwidth, angle=0]{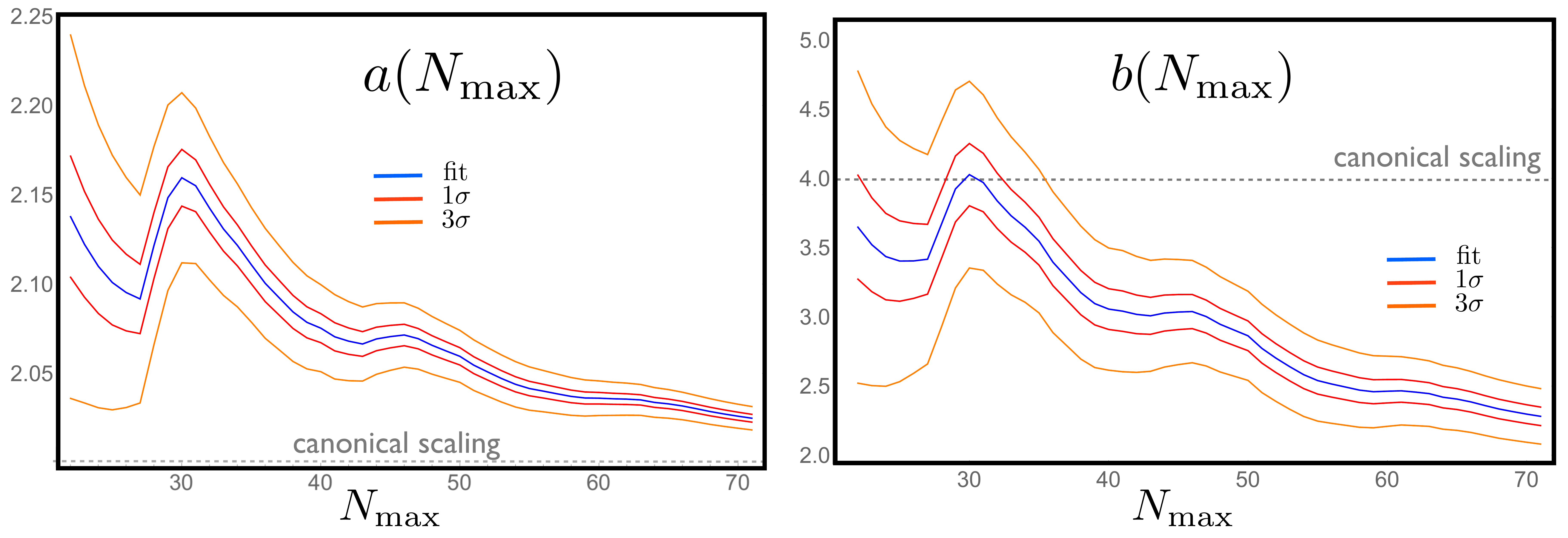}
  \end{center}
\vskip-.5cm
\caption{Approach towards Gaussian scaling. The  parameter $a$ (left panel) and $b$ (right) from \eq{eq:linear_fit_eigenvalues} are plotted  as function of $N_{\rm max}$, the highest  approximation order retained for the fit.  Shown are the fit (blue), together with the $1\sigma$- (red), and the $3\sigma$-band (orange) of variations, and the values for canonical scaling (gray dashed). While the leading coefficient seems to approach the classical limit $(a\to2)$ with increasing $N_{\rm max}$,  the subleading term $b$ increasingly deviates form its classical limit.}\label{fig:ab}
\end{figure*}
It is useful to  study the trend towards Gaussian scaling in Fig.~\ref{fig:UV} in more detail. We perform a least square fit for the large-order behaviour of scaling exponents by a simple  linear form, similar to \eq{eq:Gaussian_spectrum}. We introduce 
\begin{align}
\vartheta_n \approx a \, n - b\,, \label{eq:linear_fit_eigenvalues}
\end{align}
where $a$ and $b$ are fit parameters to be determined from the available data, \eq{theta}. In the free theory the values would read $a_G=2$ and $b_G=4$, see \eq{eq:Gaussian_spectrum}. For the fit, we use the data sets from approximation order $N=12$ to $N=N_{\rm max} = 71$, following \cite{Falls:2013bv}. We always exclude the largest two eigenvalues of each set to avoid truncation artefacts (the highest eigenvalues have, typically, not yet converged well to their asymptotic values). We find
\begin{equation}\label{fit71}
\begin{array}{rl}
a &= 2.042 \pm 0.002\\[.5ex]
b &= 2.91 \pm 0.05 \,.
\end{array}
\eeq
The slope $a$ is very close to the Gaussian slope whereas the constant $b$ exhibits a  deviation from its Gaussian value. This can also be understood from Fig.~\ref{fig:UV} where the vast majority of the data points lies above the Gaussian line. This implies that the  eigenvalues $\vartheta_n$, for sufficiently large $n$, deviate from their canonical values \eqref{eq:Gaussian_spectrum} by a correction of order $1/n$. 
We note that the fit errors only reflect the statistical fluctuations of the underlying data set. They do not account for systematical errors due to the truncation of the system. Therefore the fit values in principle can change outside the statistical error bars when the maximum approximation order $N_{\rm max}$ is varied.
Let us compare \eq{fit71} with a similar fit carried out in \cite{Falls:2013bv,Falls:2014tra} for fixed point data up until $N_{\rm max}=35$, also based on a slightly different RG flow. There the values $a= 2.17 \pm 5\%$ and $b =4.06 \pm 10 \%$ were found. We observe that  doubling the approximation order moves $a$ closer to its Gaussian value. In turn, the subleading coefficient $b$ has moved further away from Gaussian values. 

This trend can be understood by investigating how the parameters $a$ and $b$ depend on the maximal order of the  polynomial approximation, $N_{\rm max}$, retained in the fit. This is illustrated in Fig.~\ref{fig:ab}.  
The left panel shows the fit for $a(N_{\rm max})$ for different values of $N_{\rm max}$ within the interval $[22,71]$. Besides the fit itself (blue line), we also display the $1\sigma$ (red) and the $3\sigma$ (orange) bands of variation.  The smallness of the variations documents that the fit is matched by most of the data points, with only mild fluctuations overall. Moreover, the plot clearly documents the trend that $a(N_{\rm max})$ slowly decreases, possibly asymptoting towards its canonical value for $N_{\rm max}\to\infty$. Similarly, the right panel of  Fig.~\ref{fig:ab} displays the same analysis for $b(N_{\rm max})$. Here, however, we observe that $b$ increasingly deviates from Gaussian values with increasing $N_{\rm max}$. We expect both $a$ and $b$ to asymptotically approach a constant in the limit $N_{\rm max} \to \infty$. For $a$ we have indications that the limiting value may coincide with its Gaussian value, but not so for $b$.

\begin{figure}[t]
    \begin{center}
    \includegraphics[width=.7\textwidth, angle=0]{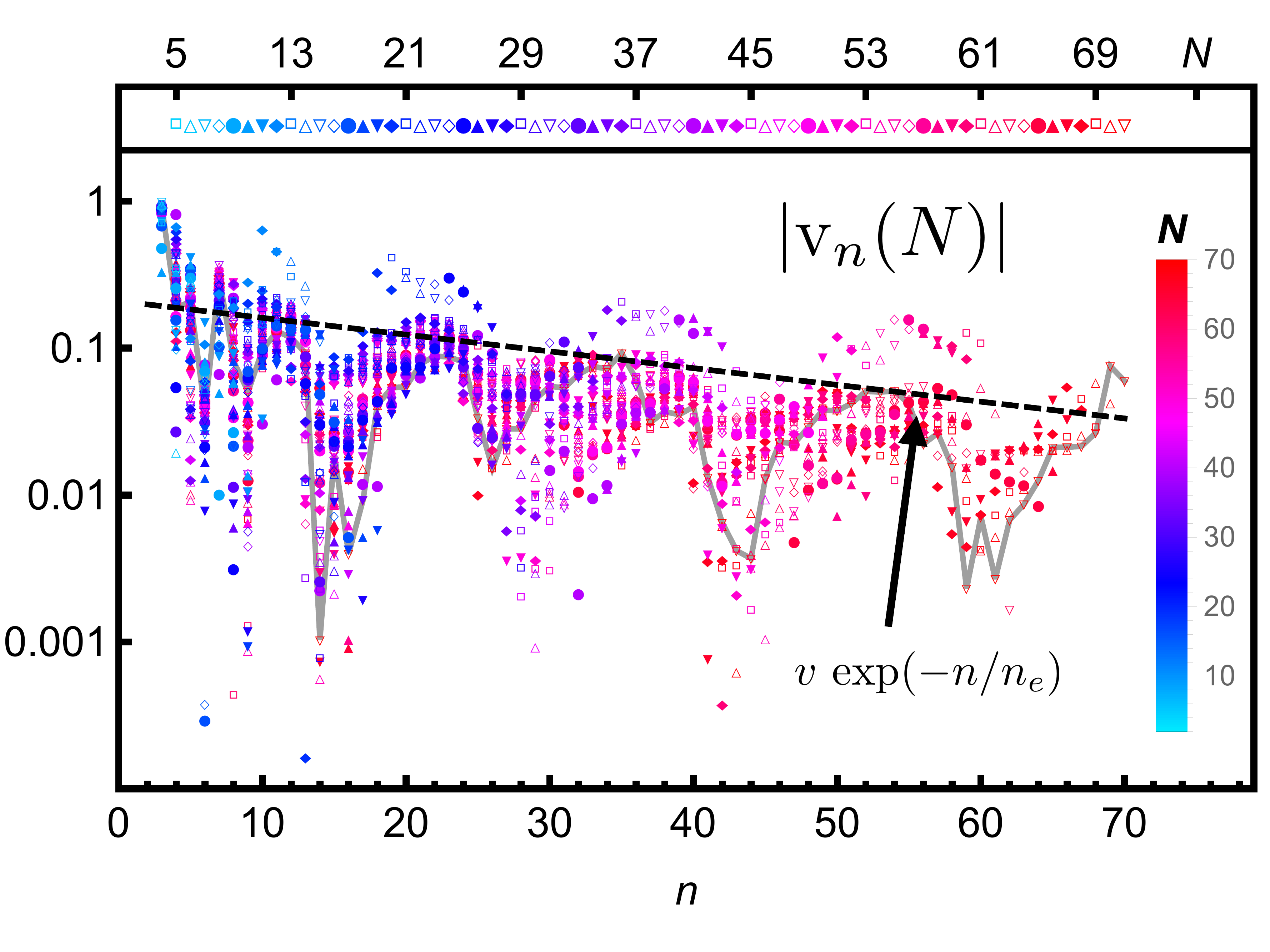}
     \end{center}
\vskip-.5cm
\caption{Signatures for weak residual interactions.
Shown are the parameters  ${\rm v}_n$ which serve as a measure for the effective coupling strengths of higher order curvature invariants,
 for all $n\ge 4$ and all approximation orders; see \eq{v}.   
The inset and the top legend explain the colour-coding of approximation orders $N$. 
The black dashed line corresponds to  a simple  exponential fit $\bar{\rm v}_n$, detailed in \eq{vbar}.
To guide the eye, the thin gray line connects data from the highest approximation. }\label{fig:NearGauss}
\end{figure}

\subsection{Fingerprints of weak coupling}\label{perturbative}

It is interesting to discuss the  approach towards Gaussian scaling from the point of view of the underlying interaction strengths.
It is well-known that weakly coupled quantum field theories with near-Gaussian fixed points \cite{Litim:2014uca,
Bond:2017sem,Buyukbese:2017ehm,Bond:2017tbw,Bond:2017wut,Bond:2017suy,Bond:2017lnq} or with exactly marginal interactions~\cite{Litim:2011bf,Heilmann:2012yf,Marchais:2017jqc,Litim:2018pxe} give rise to small deviations from Gaussian scaling.
In this light, we quantify the impact of quantum fluctuations by  measuring the deviation of scaling exponents  from classical scaling. Naturally, strong quantum effects relate to large deviations from classical scaling, and vice versa.
A virtue of this implicit definition for  ``effective coupling strengths'' is that it relates to universal quantities rather than to non-universal values of the underlying  couplings.

Following \cite{Falls:2013bv,Falls:2014tra}, we introduce the deviations from canonical scaling
by the relative distance of the corresponding universal eigenvalue $\vartheta_n$ from  Gaussian values $\vartheta_{G,n}$, 
\beq\label{v}
{\rm v}_n(N) = 1 - \frac{{\rm Re}\, \vartheta_n(N)}{\vartheta_{G,n}}\,.
\eeq
The effective couplings \eq{v}  may be expressed explicitly in terms of the polynomial couplings of  \eq{eq:f_expansion} at the fixed point. By definition, we have that  ${\rm v}_n\to 0$ for $\vartheta_n \to\vartheta_{G,n }$, which is why they can be  taken as a measure for the  strength of residual interactions. 

In Fig.~\ref{fig:NearGauss} we show our data for  ${\rm v}_n$ for all approximation orders $N$.  The inset explains the colour-coding, and the top legend details the symbols used to denote the different approximation orders $N$. The thin gray line connects data from the highest approximation order. The overall trend shows that the effective coupling strengths \eq{v} decrease with increasing $n$.  To capture this behaviour more quantatively  we consider averages of  ${\rm v}_n(N)$ over various approximation orders $N$, which we  denote by $ \bar{{\rm v}}_n$, and then search for a suitable numerical fit. A good fit is provided by a simple exponential,
 \beq\label{vbar}
 \bar{{\rm v}}_n \approx v \exp \left( -\frac{n}{n_e} \right)
 \eeq  
where $n_e$ and $v$ are two fit parameters. In Fig.~\ref{fig:NearGauss}, the black dashed line corresponds to  the fit \eq{vbar}.
 Quantitatively we find the parameters
$v = 0.209$
and
$n_e = 38.05$.
The result implies that $ \bar{{\rm v}}_n$ decreases by an order of magnitude whenever $n$ increase by $\Delta n \approx 89$. 
The important piece of information here is that the data shows a consistent, albeit slow, asymptotic decay towards near-Gaussian values.  Extrapolation of \eq{vbar} to higher orders  predicts that for sufficiently large approximation order (with $N \gg n)$ we have 
\beq
\label{v0}
{\rm v}_n(N)\to 0
\eeq
in the limit $1/n\to 0$.
This result \eq{vbar}, \eq{v0} is quite remarkable from the viewpoint of the underlying interactions. 
For a strongly coupled theory, one would have expected $O(1)$ deviations from classical scaling for any of the universal eigenperturbations, while a weakly coupled theory would give minute deviations only. In the model of quantum gravity considered here, we face a mixture of either of these: for the few relevant eigenperturbations, $i.e.$~those due to the canonically relevant and marginal couplings, the quantum corrections are strong and modify scaling exponents by order unity. In turn, for almost all irrelevant eigenperturbations, $i.e.$~those due to couplings with increasingly large canonical mass dimension, the quantum effects on universal eigenvalues are increasingly minute, despite of the fact that the fluctuations are non-trivial and that higher order vertex functions remain non-zero.\footnote{See \cite{Eichhorn:2018ydy} for  signatures of weak coupling within Einstein-Hilbert-matter systems.}

As a closing remark, we stress  that near-Gaussian behaviour such as \eq{vbar} and \eq{v0} is not mandatory for asymptotic safety. Rather, the much weaker constraints ${\rm v}_n(N)<1$ are still in accord with the principles of asymptotic safety and a finite dimensional UV critical surface. 
From the viewpoint of the full theory for quantum gravity, our  findings  thus consolidate the results  \cite{Falls:2013bv,Falls:2014tra,Falls:2017lst} that ``most of quantum gravity'' is rather weakly coupled, except for a few dominant interactions which relate predominantly to the vacuum energy, Newton's coupling, and a fourth order interaction. Equally important is the  observation  that   higher-order interactions help to anchor the fixed point, yet their precise form does not seem to matter substantially \cite{Falls:2017lst}. This pattern of results has interesting implications for ``quantum gravity model building''. Most notably,  it offers guidance for the construction of   simplified models capturing the key features of the full theory. It will be  useful to test and exploit the signatures of weak coupling  within  alternative approaches to quantum gravity such as the lattice,  loops, strings, or other.

\subsection{de Sitter solutions from quantum gravity}
\label{Inflation}

Finally, we discuss  implications of quantum gravity fixed points  for  (quantum) cosmology and phases of inflationary expansion. Of particular  interest is the existence or not of de Sitter solutions in the quantum regime of the theory  \cite{Hindmarsh:2011hx,Falls:2016wsa}.
There are strong observational indications for inflationary phases  both during the early-time \cite{Ade:2013uln,Ade:2013zuv} and the late-time cosmological evolution of the  universe \cite{Perlmutter:1998np}.  Inflation may  well be generated quantum-gravitationally provided the scalar curvature $R$ of the quantum effective action becomes nearly constant so that the cosmological evolution becomes  similar to that of a de Sitter universe. The real positive solutions of the quantum equation of motion $E(R)=0$ with
$E(R)=2 f(R)-R f'(R) $
then determine the corresponding Hubble parameter. Here we will concern ourselves with saddle points of Euclidean action with constant curvature. Under the assumption that these provide an appropriate background around which a Wick rotation can be performed we tentatively interpret these as corresponding to Lorentzian spacetimes with the same curvature. We assume therefore that four-spheres are analytically continued to de~Sitter space and  hyperbolic spaces of constant negative curvature are analytically continued to anti-de~Sitter space. 

In asymptotically safe versions of gravity, (anti-)de Sitter solutions  arise by default in the Einstein Hilbert approximation as soon as the  cosmological constant comes out positive (negative) in the UV. Also, de Sitter solutions may arise along UV-IR connecting trajectories provided the RG scale is suitably identified with, $e.g.$, the inverse cosmological time  $(k\propto 1/t$) \cite{Bonanno:2001xi},   Ricci curvature $(k^2\propto R$) \cite{Hindmarsh:2012rc}, or through Bianchi identities \cite{Hindmarsh:2011hx}.
Results for approximations beyond Einstein-Hilbert have also been found.  
Within $f(R)$  quantum gravity,
it was investigated in \cite{Bonanno:2010bt} whether polynomial expansions 
of the flow  \eq{flowtraces_I} display de Sitter solutions.  High-order studies 
and resummations have  established that there are none for the flow  \eq{flowtraces_I}, albeit narrowly, and up to background Ricci curvature of the order of RG scale \cite{Falls:2016wsa}.
Instead, 
complex conjugate pair of de Sitter solutions were found order by order, with a small imaginary part (dubbed ``near de Sitter''), alongside an anti-de Sitter solution.  It was also noticed that small variations of the fixed point couplings may generate de Sitter solutions  \cite{Falls:2016wsa}. 
A single de Sitter solution has  also been reported 
within  an  exponential parameterisation of the metric fluctuations \cite{Ohta:2015efa,Ohta:2015fcu} and  within a geometric formulation of the flow \cite{Demmel:2015oqa}. 
In \cite{Christiansen:2017bsy} a vertex expansion around a spherical background was performed and the equation of motion has been calculated from the flow of the background effective action as well as from the one-point function. While no solution was found in the former, an anti-de Sitter solution does appear in the latter. Finally, in  \cite{Falls:2017lst} it was found that high-order polynomial actions also  involving Ricci tensor interactions $R_{\mu\nu}R^{\mu\nu}$ and powers thereof equally generate an AdS and two dS solutions in the fixed point regime.

\begin{figure*}[t]
    \begin{center}
    \includegraphics[width=.7\textwidth, angle=0]{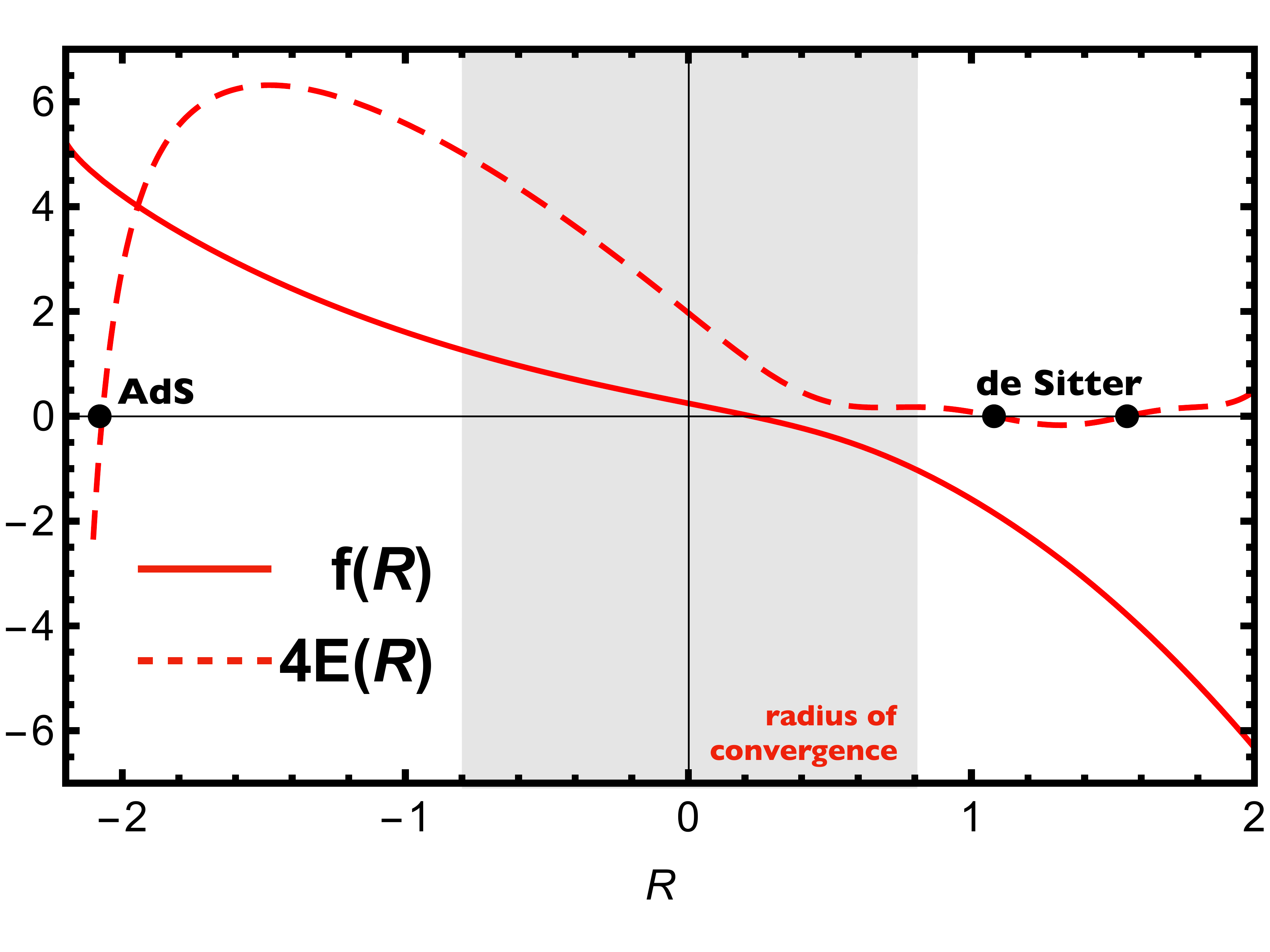}
 \end{center}
\vskip-.5cm
\caption{Fixed point action and quantum equations of motion from numerical integration and Pad\'e\ resummation. Shown are the functions $f(R)$ (full  line) and $4E(R)$ (dashed line)   within their full domains of validity $|R|<R_+$. We observe that  $f'(R)<0$ throughout. The quantum equations of motion offers two de Sitter and an anti de Sitter solution \eq{dS}  (black dots), just outside the polynomial radius of convergence \eq{eq:radius_of_convergence_pure_gravity} (gray-shaded area).}
\label{fig:fE}
\end{figure*}

In Fig.~\ref{fig:fE} we display our new findings for the scaling solution $f(R)$ (full line) and for the quantum equation of motion (dashed line) from the new flow equation \eq{flowtraces}. The gray-shaded area indicates the polynomial radius of convergence \eq{eq:radius_of_convergence_pure_gravity}. Results in the full domain of validity  up to $|R|<R_+$ \eq{poles} are obtained by numerical integration of \eq{eq:fixed_point}.
Overall, we find that gravity remains anti-screening for all background curvatures $(f'(R)<0)$. 
We also observe that de Sitter solutions are absent within the polynomial radius of convergence, in agreement with earlier findings \cite{Falls:2016wsa}.
For larger Ricci curvature of the order of the RG scale, however, and within the range \eq{poles}, two de Sitter solutions and an anti de Sitter solution are found 
\beq \label{dS}
\begin{array}{lcr}
R^{\rm AdS}_-&=&-2.071\,,\\[1ex]
R^{\rm dS}_+&=&1.080\,, \quad\\[1ex]
R^{\rm dS}_{++}&=&1.548\,.
 \end{array}
\eeq
Results are confirmed independently with  Pad\'e resummations of the polynomial fixed point solution. While the AdS solution  is qualitatively in agreement with  \cite{Falls:2016wsa}, the interesting novelty is the occurrence of two de Sitter solutions. The difference is  traced back to the improved flow  \eq{flowtraces} and the absence of technical poles \eq{pole}. Albeit numerically small, these differences turn the complex conjugate de Sitter solutions of the flow \eq{flowtraces_I}   \cite{Falls:2016wsa} into two real de Sitter solutions \eq{dS}.  
With all other approximations being equal, we conclude that the removal of  technical artifacts around $R\approx 3$ and  $R\approx 4$ has enabled real de Sitter solutions in the fixed point regime.  In this light, the considerations of \cite{Bonanno:2010bt} can straightforwardly be applied

It is worth comparing our results \eq{dS} with those from the quantum gravity model  of \cite{Falls:2017lst}, where
 the higher curvature terms beyond Einstein-Hilbert are of the form $(R_{\mu\nu}R^{\mu\nu})^n$ and $\propto R\cdot (R_{\mu\nu}R^{\mu\nu})^n$ for $n\ge 1$, and thus quite different from the powers of the Ricci scalar $\propto R^{n+1}$  employed in this work.  The vacuum solutions  of \cite{Falls:2017lst} is given by
\beq \label{Ricci-dS}
\begin{array}{lcr}
R^{\rm AdS}_-&=&-2.143\,,\\[1ex]
R^{\rm dS}_+&=&0.996\,, \quad\\[1ex]
R^{\rm dS}_{++}&=&1.458\,. \quad
 \end{array}
\eeq
The results \eq{Ricci-dS} are very close to those found here, \eq{dS}, despite of the differences in the underlying models. The quantitative similarity would seem to suggest that the existence of de Sitter solutions such as \eq{dS} or \eq{Ricci-dS} is a genuine feature, potentially  independent of the type of higher curvature terms retained in the effective action. 
However, the vacuum solutions \eq{dS} are non-perturbative in that they arise for curvatures  {\it outside} the polynomial radius of convergence \eq{eq:radius_of_convergence_pure_gravity}. As such, the de Sitter solutions become visible only after numerical integration beyond the polynomial radius of convergence, or after Pad\'e resummation. This aspect is different from  the de Sitter solutions \eq{Ricci-dS} since the latter arise strictly {\it within} the radius of convergence of the corresponding model \cite{Falls:2017lst}, which came out more than twice as big as the radius found here, \eq{eq:radius_of_convergence_pure_gravity}.  In either setting, the de Sitter solutions arise for curvature still within the range \eq{poles} dictated by the fluctuations-induced poles of the flow.
Also, reducing either setting to the Einstein-Hilbert action would  yield a single de Sitter solutions with a
 curvature roughly half as large as \eq{dS}, see Fig.~\ref{fig:fE}.

The results \eq{dS} and \eq{Ricci-dS} with two different types of de Sitter solutions in the UV  is  noteworthy for phases with accelerated expansion in cosmology \cite{Falls:2016wsa,Falls:2017lst}.    
Specifically, it suggests a scenario of (quantum) cosmology in which two distinct vacua are already available in the quantum gravity domain at the UV fixed point. If these two solutions continue to persist along RG trajectories running out of the UV fixed point towards lower energies, they allow for separate phases of accelerated cosmological expansions. Further studies that analyse  the flows away from the fixed point should indicate whether such trajectories exist.  
The first one would end after a finite cosmological time with $R_{\rm dS}= R_{\rm infl}$ and $f(R) \propto R^2_{\rm infl}$ leading to inflation in the early universe \cite{Starobinsky:1980te}. The second one for which $R_{\rm dS} = 4 \Lambda$ (where $\Lambda$ is the cosmological constant) gives rise to the late-time
accelerated expansion as soon as the contributions to the energy density of the universe are dominated by $\Lambda$. 
 We conclude that early and late-time phases of accelerated cosmological expansion may very well originate from the quantum  gravity.

\section{\bf Discussion and conclusions} \label{Conclusion}
We have investigated the asymptotic safety conjecture for models of  quantum gravity using Wilson's renormalisation group. The main novelty are improved momentum cutoffs,  chosen in such a way that technical curvature poles along the renormalisation group flow 
are absent  \eq{flowtraces}, \eq{eq:fixed_point}. A powerful algebraic code was developed  to exploit exact recursive relations determining the fixed point in polynomial approximations up to including $N=70$ powers of the Ricci scalar   $\sim \int \sqrt{g}R^{N}$
\cite{Schroeder:Thesis},  
doubling-up over previous efforts \cite{Falls:2013bv,Falls:2014tra,Falls:2016wsa}.
The existence of a stable UV fixed point (Tab.~\ref{tCouplings}) is confirmed  with well-converging  scaling exponents, and  in accord with expectations guided by canonical power counting (Figs.~\ref{fig:couplings},~\ref{fig:eigenvalues_bootstrap}). Predictivity in the UV --- which lies at the heart of the asymptotic safety conjecture --- is associated with a finite number of relevant couplings. The corresponding UV critical surface is confirmed to be three-dimensional (Figs.~\ref{fig:UVsurface},~\ref{fig:UV}), again in good  agreement with earlier findings. We conclude that the fixed point and its universal aspects are  stable and only mildly modified by the removal of spurious poles. 

From the viewpoint of the full quantum theory of gravity beyond the approximations adopted here, we have argued that our results offer the lower bound \eq{free} on the number of fundamentally free parameters. In addition, our findings  suggest 
that an  upper bound is approximately given by the number of canonically relevant and marginal couplings of the theory.
We have
exploited the validity of canonical power counting \cite{Falls:2013bv,Falls:2014tra} for which abundant evidence has been given (Figs.~\ref{fig:eigenvalues_bootstrap},~\ref{fig:UV}). 
We  have also discussed  mild tensions amongst earlier findings and explained why  differences in the underlying projection methods can lead to  qualitative differences in the eigenvalue spectrum. This is particularly relevant  for the canonically marginal interactions where    small quantum effects may tip their relevancy  either way.

Another important result is the confirmation of near-Gaussian scaling (Fig.~\ref{fig:ab}).
A finite number of relevant eigenvalues  (Fig.~\ref{fig:UVsurface}) would have already been sufficient for asymptotic safety to arise.   
Instead, and despite of the interacting nature of the fixed point,  the universal eigenvalues \eq{eq:linear_fit_eigenvalues} 
approach near-classical values \eq{eq:Gaussian_spectrum} with increasing canonical mass dimension, modulo mild subleading corrections.
A similar pattern of findings is known from exactly solvable non-gravitational models 
where  near-canonical scaling
is traced back to near-Gaussian fixed points \cite{Litim:2014uca,
Bond:2017sem,Buyukbese:2017ehm,Bond:2017tbw,Bond:2017wut,Bond:2017suy,Bond:2017lnq} or the occurrence of strictly marginal interactions~\cite{Litim:2011bf,Heilmann:2012yf,Marchais:2017jqc,Litim:2018pxe}. 
Deviations  of irrelevant eigenperturbations from canonical scaling \eq{v}, \eq{vbar}
can then be taken as small parameters in  the sense of perturbation theory. 
Our findings thus  consolidate 
that ``most of quantum gravity'' is rather weakly coupled (Fig.~\ref{fig:NearGauss}), except for a few dominant interactions \cite{Falls:2013bv,Falls:2014tra,Falls:2017lst}. It is also confirmed that the precise form of higher-order interactions is of subleading relevance for this pattern \cite{Falls:2017lst}. Our  results have
interesting  conceptual and practical implications for ``quantum gravity model building'', and, possibly, for other approaches to quantum gravity.

A qualitatively new result has arisen in the context of cosmology. The quantum equation of motion show the existence of de Sitter solutions (Fig.~\ref{fig:fE}), a prerequisite for cosmological scaling solutions. Here, de Sitter solutions arise as a direct consequence of the absence of
spurious 
curvature poles. Earlier studies have  found near de Sitter solutions in an otherwise identical set-up \cite{Falls:2016wsa}. Our findings seem to illustrate how important it can be  to avoid technical artefacts of the RG flow even if their effects are quantitatively moderate. The vacuum solutions compare well with those found
recently in models with additional  Ricci tensor invariants  \cite{Falls:2017lst}. It will be interesting to see whether de Sitter solutions persist in even more sophisticated models of asymptotically safe quantum gravity.  
The presence of two de Sitter solutions opens a door for quantum gravity-induced models of inflationary expansions at early or late cosmological times, and comparison with data \cite{Ade:2013uln,Ade:2013zuv,Perlmutter:1998np} along the lines of \cite{Hindmarsh:2011hx}. We hope to come back to these topics in future work.

\section*{\bf Acknowledgements}
We thank Kostas Nikolakopoulos for  discussions. This work has been supported by the Studienstiftung des Deutschen Volkes (JS).  Part of this work was performed at the Aspen Center for Physics, which is supported by National Science Foundation grant PHY-1607611. DL  also gratefully acknowledges financial support by the Simons Foundation.
JS gratefully acknowledges hospitality and support by the Department of Theoretical Physics, U Heidelberg.

\bibliography{bib_Rmunu}
\bibliographystyle{JHEP}

\end{document}